\documentstyle[11pt,preprint]{aastex}

\newcommand{\PNM}{P(N|M)}
\newcommand{\PNN}{P(N|\left<N\right>)}
\newcommand{\Mmin}{M_{\mathrm{min}}}
\newcommand{\Mcrit}{M_{\mathrm{crit}}}

\newcommand{\ng}{\bar{n}_g}
\newcommand{\vh}{{\bf v}_{\mathrm{h}}}
\newcommand{\vg}{{\bf v}_{\mathrm{g}}}
\newcommand{\vm}{{\bf v}_{\mathrm{m}}}

\newcommand{\rcen}{r_{\mathrm{cen}}}
\newcommand{\rpcb}{r_{\mathrm{pcb}}}
\newcommand{\Rvir}{R_{\mathrm{vir}}}

\newcommand{\hvol}{h^{3}{\mathrm{Mpc}}^{-3}}
\newcommand{\hmpc}{h^{-1}\mathrm{Mpc}}
\newcommand{\hkpc}{h^{-1}\mathrm{kpc}}

\newcommand{\Msun}{M_{\odot}}
\newcommand{\kms}{{\,{\rm km}\,{\rm s}^{-1}}}
\newcommand{\Omegam}{\Omega_{m}}
\newcommand{\Omegab}{\Omega_{b}}
\newcommand{\Omegal}{\Omega_{\Lambda}}

\newcommand{\N}{\left<N\right>}
\newcommand{\Nsqr}{\left<N^2\right>}

\newcommand{\NN}{\left<N(N-1)\right>}
\newcommand{\NNN}{\left<N(N-1)(N-2)\right>}
\newcommand{\xig}{\xi_g(r)}

\newcommand{\Mbmin}{M_{b,\mathrm{min}}}
\newcommand{\Mmax}{M_{\mathrm{max}}}
\newcommand{\tmax}{t_{\mathrm{max}}}
\newcommand{\tstart}{t_{\mathrm{start}}}
\newcommand{\tstop}{t_{\mathrm{stop}}}
\newcommand{\xibar}{\bar{\xi}}

\bibpunct[,]{(}{)}{;}{a}{}{,}
\begin{document}

\title{The Halo Occupation Distribution and the Physics of Galaxy Formation}

\author{
Andreas A. Berlind \altaffilmark{1},
David H. Weinberg \altaffilmark{2},
Andrew J. Benson \altaffilmark{3},
Carlton M. Baugh \altaffilmark{4}, 
Shaun Cole \altaffilmark{4}, 
Romeel Dav\'e \altaffilmark{5}, 
Carlos S. Frenk \altaffilmark{4},
Adrian Jenkins \altaffilmark{4},
Neal Katz \altaffilmark{6},
and Cedric G. Lacey \altaffilmark{4,7}
}
\altaffiltext{1}{Center for Cosmological Physics and Department of Astronomy \& Astrophysics, The University of Chicago, Chicago, IL 60637, USA; Email: aberlind@oddjob.uchicago.edu}
\altaffiltext{2}{Department of Astronomy, The Ohio State University, 
Columbus, OH 43210, USA; Email: dhw@astronomy.ohio-state.edu}
\altaffiltext{3}{Department of Astronomy, California Institute of Technology, 
Pasadena, CA 91125, USA; Email: abenson@astro.caltech.edu}
\altaffiltext{4}{Institute for Computational Cosmology, University of Durham, Durham DH1 3LE, UK; Email: c.m.baugh, Shaun.Cole, c.s.frenk, A.R.Jenkins@durham.ac.uk}
\altaffiltext{5}{Steward Observatory, University of Arizona, Tucson, AZ 85721, USA; Email: rad@as.arizona.edu}
\altaffiltext{6}{Department of Physics and Astronomy, University of Massachusetts, 
Amherst, MA 01003, USA; Email: nsk@kaka.phast.umass.edu}
\altaffiltext{7}{Observatoire de Lyon, 9 Avenue Charles Andre, 69230 Saint Genis Laval, France; Email: lacey@obs.univ-lyon1.fr}

\begin{abstract}
The halo occupation distribution (HOD) describes the bias between galaxies and
dark matter by specifying (a) the probability $\PNM$ that a halo of virial mass
$M$ contains $N$ galaxies of a particular class and (b) the relative spatial
and velocity distributions of galaxies and dark matter within halos.  We
calculate and compare the HODs predicted by a smoothed particle hydrodynamics
(SPH) simulation of a $\Lambda$CDM cosmological model (cold dark matter with a
cosmological constant) and by a semi-analytic galaxy formation model applied to
the same cosmology. Although the two methods predict different galaxy mass
functions, their HOD predictions for samples of the same space density agree
remarkably well. In a sample defined by a baryonic mass threshold, the mean
occupation function $\N_M$ exhibits a sharp cutoff at low halo masses, a slowly
rising plateau in which $\N$ climbs from one to two over nearly a decade in
halo mass, and a more steeply rising, high occupancy regime at high halo mass.
In the low occupancy regime, the factorial moments $\NN$ and $\NNN$ are well
below the values $\N^2$ and $\N^3$ expected for Poisson statistics, with
important consequences for the small scale behavior of the 2- and 3-point
correlation functions. The HOD depends strongly on galaxy age, with high mass
halos populated mainly by old galaxies and low mass halos by young galaxies.
The distribution of galaxies within SPH halos supports the assumptions usually
made in semi-analytic calculations: the most massive galaxy lies close to the
halo center and moves near the halo's mean velocity, while the remaining,
satellite galaxies have the same radial profile and velocity dispersion as the
dark matter. The mean occupation at fixed halo mass in the SPH simulation is
independent of the halo's larger scale environment, supporting both the
merger tree approach of the semi-analytic method and the claim that the HOD
provides a complete statistical characterization of galaxy bias. We discuss
the connections between the predicted HODs and the galaxy formation physics
incorporated in the SPH and semi-analytic approaches. These predictions offer
useful guidance to theoretical models of galaxy clustering, and they will be
tested empirically by ongoing analyses of galaxy redshift surveys. By applying
the HODs to a large volume N-body simulation, we show that both methods predict
slight departures from a power-law galaxy correlation function, similar to 
features detected in recent observational analyses.
\end{abstract}



\section{Introduction} \label{intro}

A complete theory of galaxy formation should predict the distributions
of galaxy luminosities, colors, sizes, and morphologies, the correlations
among these properties, and the relation between the spatial clustering
of any given class of galaxies and that of the underlying
dark matter distribution.  
This last class of predictions, the ``bias''
of galaxies as a function of their observable properties, 
is becoming an increasingly important test of theoretical models
thanks to the new generation of large galaxy redshift surveys, 
in particular the 2dF Galaxy Redshift Survey (2dFGRS; \citealt{colless01})
and the Sloan Digital Sky Survey (SDSS; \citealt{york00}).
The ``halo occupation distribution'' (HOD) formalism 
is an especially powerful framework for carrying out such tests;
it characterizes the bias of a class of galaxies by the probability $\PNM$ 
that a halo of virial mass $M$ contains $N$ such galaxies and additional
prescriptions that specify the relative distributions of galaxies
and dark matter within halos.
If the HOD at fixed halo mass is statistically independent of the
halo's large scale environment, as theoretical models predict
(\citealt{bond91,white96,lemson99}; this paper), then this
description of galaxy bias is essentially complete: given the HOD
and the halo population predicted by a particular cosmological model,
one can calculate any galaxy clustering statistic, on scales from the
linear regime to the deeply non-linear regime.
Empirical determinations of the HOD for different galaxy types
would therefore summarize everything that observed galaxy clustering
has to say about the physics of galaxy formation, in a form that can
be readily compared to theoretical predictions.

This paper examines the HODs predicted by the two leading theoretical
methods for studying galaxy formation and bias in a cosmological context:
semi-analytic models (e.g., 
\citealt{white91,kauffmann93,cole94,avila98,somerville99})
and hydrodynamic numerical simulations (e.g.,
\citealt{cen92,katz92,evrard94,pearce99,white01,yoshikawa01}).
We apply both methods to a $\Lambda$CDM cosmological model (inflationary cold
dark matter with a cosmological constant), adopting the same cosmological
parameters in each case.  We focus on galaxy samples defined by 
thresholds in baryon mass and stellar population age, which
are roughly analogous to observational samples defined by cuts in
luminosity and color.  
Except for using the same cosmological model and thus the same
present-day halo population, we do not make any efforts to ``tune''
the semi-analytic calculation to match the hydrodynamic simulation;
we apply each method in its ``standard'' form.
We present results for the various features
of the HOD --- the mean occupation $\N$ as a function of halo mass,
moments of the distribution $\PNN$, and the spatial and velocity
distributions of galaxies within halos --- and
we interpret these features in terms of the physical 
processes represented in the theoretical models.
In the short term, our results should provide useful input to
theoretical models of galaxy clustering by allowing the predictions
of these galaxy formation models to be ``bootstrapped'' into analytic
calculations or larger volume N-body simulations, and they offer guidance
to efforts to infer parameters of the HOD from observational data.
In the slightly longer term, these predictions will be
tested by empirical determinations of the HOD, and any discrepancies
with observations may point the way to necessary revisions
of the galaxy formation model or the underlying cosmological model.

Several aspects of the HOD predicted by semi-analytic models have been 
investigated in the pioneering papers of \cite{kauffmann97},
\cite{governato98}, \cite{kauffmann99}, and \cite{benson00}, which used 
semi-analytic methods to assign galaxy populations to the halos of 
N-body simulations.  \cite{seljak00} and \cite{sheth01a} measured 
$\PNM$ from the \cite{kauffmann99} models and incorporated them into 
analytic predictions of galaxy clustering and bias (see also
\citealt{scoccimarro01,sheth01,scranton02}).
HOD predictions of hydrodynamic simulations have been presented
by \cite{white01} at redshifts $z=3$ and $z=1$ and by 
\cite{yoshikawa01} at $z=3$, 2, and 0.
Relative to this earlier work, our examination of the HOD in this
paper is more comprehensive, and our side-by-side comparison of
numerical and semi-analytic results for the same cosmological
model allows us to evaluate the robustness and limitations of
the predictions and to better understand the physics that gives 
rise to them.

Other studies of the halo occupation distribution have focused on the 
connections between the HOD and statistical measures of galaxy clustering.
Many of these studies have utilized the power of the HOD formalism
and the related ``halo model'' of dark matter clustering as 
a tool for analytic calculations 
\citep{ma00,seljak00,benson01,scoccimarro01,sheth01,white01b,cooray02b},
drawing on methods developed over the course of several decades
\citep{neyman52,peebles74,mcclelland77,scherrer91,mo96,sheth01b}.
Berlind \& Weinberg (\citeyear{berlind02}, hereafter BW) computed
the impact of HOD bias on many of the most widely used galaxy
clustering statistics by applying parameterized HOD models to an N-body 
simulation of the $\Lambda$CDM scenario.  They concluded that different
statistics constrain the HOD in complementary ways, making it
possible to determine the HOD empirically from observed galaxy clustering,
at least for a known cosmological model.  Important steps toward
observational determination of the HOD of bright, optically selected
galaxies, drawing mainly on the 2- and 3-point correlation functions, 
the group multiplicity function, and galaxy-galaxy lensing, have been taken 
by \cite{jing98}, \cite{peacock00}, \cite{scoccimarro01}, \cite{guzik02}, 
\cite{marinoni02}, \cite{yang03}, \cite{vandenbosch02}, \cite{zehavi03}, 
and \cite{magliocchetti03}.  \cite{kochanek03}, 
\cite{jing02}, and \cite{cooray02} have investigated HOD constraints for 
galaxies selected in the near- or far-infrared, and \cite{wechsler01}, 
\cite{bullock02}, and \cite{moustakas02} have applied similar methods to 
high-redshift galaxies.
An ambitious but, we think, realizable goal is to use the high-precision
measurements afforded by the 2dFGRS and the SDSS 
to break the ``degeneracies'' between cosmology and bias,
obtaining tight, simultaneous constraints on the mass function
and clustering of dark halos 
and the HODs of many different galaxy classes (see discussions by BW,
\citealt{zheng02}, and \citealt{weinberg02b}).
We hope that the results presented here will provide inspiration
to such efforts, by illustrating how measurements of the HOD can
test basic ideas about the physics of galaxy formation.

Our approach to the HOD is essentially the one described by BW,
which was in turn inspired largely by the discussion of 
Benson et al. (2000ab).
The clustering of galaxies predicted by the semi-analytic model
and hydrodynamic simulation investigated here, as quantified by
more traditional statistics, has been presented in separate papers
\citep{benson00,weinberg02a}.
We briefly describe the two calculations and the selection of galaxy 
populations in \S~\ref{models}.
In \S~\ref{pnm} we compare the predictions of $\PNM$ for several
different galaxy classes, and we demonstrate that $\PNM$ predicted
by the hydrodynamic simulation is independent of large scale environment,
to within the statistical limitations of our measurements.
Semi-analytic models do not predict the distribution of galaxies
within halos, but clustering calculations based on these models
usually assume that each halo contains one central galaxy moving
at the halo's center-of-mass velocity and that other ``satellite''
galaxies trace the halo's dark matter distribution.
In \S~\ref{profile}, we show that these assumptions hold to a good
approximation in the hydrodynamic simulation, which predicts
the galaxy positions and velocities directly.
In \S~\ref{central} we compare the properties of central and satellite 
galaxies in the two methods.  
The precision in predictions of the galaxy correlation function
by these two models has been limited by 
the finite size of the simulation volumes
in which they were implemented.  
In \S~\ref{xiJen} we combine the HOD results with a new, large-volume
N-body simulation to make improved predictions for the galaxy correlation
function, focusing on predicted departures from a power-law form.
In \S~\ref{conclusions} we summarize our results and discuss what
they tell us about the physical factors that shape the HOD
and thereby determine the bias between galaxies and dark matter.


\section{Theoretical Models} \label{models}

\subsection{SPH Simulation} \label{models:sph}

We use a smoothed particle hydrodynamics (SPH) simulation of a $\Lambda$CDM 
cosmological model, with $\Omegam=0.4$, $\Omegal=0.6$, $\Omegab=0.02h^{-2}$,
$h\equiv H_0/(100~\mathrm{km~s}^{-1}~Mpc^{-1})=0.65$, $n=0.95$, and 
$\sigma_8=0.8$.  This model is in good agreement with a wide variety of 
cosmological observations (see, e.g., \citealt{spergel03}), though the value of
$\Omegam$ is somewhat higher than favored by the most recent constraints.
In terms of HOD predictions, lowering $\Omegam$ while keeping other parameters 
fixed would primarily shift the mass scale of halos by a constant factor 
\citep{zheng02}, though the change of dynamical growth timescales relative 
to gas cooling timescales could have a secondary influence.
The simulation uses the Parallel TreeSPH code \citep{hernquist89,katz96,dave97} 
to follow the evolution of $144^3$ gas and $144^3$ dark matter particles 
in a $50\hmpc$ box from $z=49$ to $z=0$.  The mass of each dark matter 
particle is $6.3\times10^9 M_{\odot}$, the mass of each baryonic particle is 
$8.5\times10^8 M_{\odot}$, and the gravitational force softening is
$\epsilon_{\rm grav}=7\hkpc$ (Plummer equivalent).
As a test of numerical resolution effects, in \S\ref{pnm} we also show
results from a simulation of a $22.222\hmpc$ cube with a factor of eight
higher mass resolution ($2\times 128^3$ particles) and a gravitational
softening of $3.5\hkpc$.

Dark matter particles are only affected by gravity, whereas gas particles are 
subject to pressure gradients and shocks, in addition to gravitational forces.  
The TreeSPH code includes the effects of both radiative and Compton cooling. 
TreeSPH also includes heating by a background UV radiation field but we
only include its effects in the simulation of the $22.222\hmpc$ cube.  We
cannot accurately include a background UV radiation field at the
lower resolution of the $50\hmpc$ simulation \citep{weinberg97}.  Star 
formation is assumed to happen in regions that are Jeans unstable and where the 
gas density is greater than a threshold value ($n_H \geq 0.1 {\rm cm}^{-3}$)
and colder than a threshold temperature ($T \leq 30,000\,$K).
Once gas is eligible to form stars, it does so at a rate 
proportional to $\rho_{\mathrm{gas}}/t_{\mathrm{gas}}$, where 
$\rho_{\mathrm{gas}}$ is the gas density and $t_{\mathrm{gas}}$ is the longer 
of the gas cooling and dynamical times.  Gas that turns into stars becomes 
collisionless and releases energy back into the surrounding gas via supernova 
explosions.  A Miller-Scalo (\citeyear{miller79}) initial mass function of 
stars is assumed, and stars of mass greater than $8M_{\odot}$ become 
supernovae and inject $10^{51}$ergs of pure thermal energy into neighboring 
gas particles.  The star formation and feedback algorithms are discussed 
extensively by \cite{katz96}, and the particular simulations employed here are
described in greater detail by \cite{murali02}, \cite{dave02}, and
\cite{weinberg02a}.  The parameters are all chosen on the basis of a priori 
theoretical and numerical considerations and are not adjusted to match any
observations.

SPH galaxies are identified at the sites of local baryonic density maxima using 
the SKID algorithm,\footnote{See
{\tt http://www-hpcc.astro.washington.edu/tools/skid.html} 
and \cite{katz96}.} which selects gravitationally bound groups of star 
and cold, dense gas particles.  Because dissipation greatly increases the 
density contrast of these baryonic components, there is essentially no 
ambiguity in the identification of galaxies.  We retain only those 
particle groups whose mass exceeds a threshold 
$\Mbmin=5.42\times10^{10} M_{\odot}$, corresponding to the mass of 64 SPH 
particles, and the resulting galaxy space density is
$\ng=0.02\hvol$.  We also construct lower density (and thus more massive) 
samples with $\ng=0.01\hvol$ and $\ng=0.005\hvol$, which have minimum baryonic 
masses of $1.25\times10^{11} M_{\odot}$ and $2.39\times10^{11} M_{\odot}$, 
respectively.  The galaxy properties that we use in this analysis, aside from 
position and velocity, are the total baryonic mass and the median stellar age 
(i.e., the look-back time to the point at which half of the stellar mass had 
formed).  

We identify dark matter halos in the mass distribution using a 
friends-of-friends algorithm \citep{davis85} with a linking length of 
0.173 times the mean inter-particle separation, and we only consider halos 
consisting of at least 32 dark matter particles.  We choose this particular 
linking length because, for this cosmological model, it most closely 
corresponds to the definition of a halo assumed in the semi-analytic model 
described in the following section.  We assume a universal baryon to matter 
ratio and scale halo dark matter masses up by a factor 
$\Omegam/(\Omegam-\Omega_b)=1.134$ to obtain total halo masses. 
The minimum halo mass we resolve is therefore $2.29\times10^{11} M_{\odot}$,
and the most massive halo in our $50\hmpc$ box is 
$3.29\times 10^{14} M_{\odot}$.  
Finally, we decide halo membership of galaxies by assigning each galaxy to the 
halo that contains the dark matter particle closest to the galaxy center-of-mass.

\subsection{Semi-Analytic Model} \label{models:sa}

Semi-analytic galaxy formation models have their roots in the work of 
\citet{white78}, \citet{fall80}, \citet{cole91}, \citet{lacey91}, and 
\citet{white91}, who established the basic framework of this approach.  
The semi-analytic (SA) model that we use in this paper is GALFORM,
which is described in detail by \citet{cole00}.  The model begins with a 
population of halo masses that is usually either generated using the 
\citet{press74} halo mass function or drawn from an N-body simulation.  In 
this study, we supply the SA model with the same halo population identified in 
the SPH simulation.  The two methods' predictions can thus be compared halo by 
halo without being subject to differences caused by sample variance.  In 
addition, we produce 10 SA realizations for each SPH halo so that we can 
determine the SA $\PNM$ relation more accurately.

For each halo, the SA model first employs a Monte Carlo method to generate a 
``merger tree'', which describes the hierarchical growth of that particular 
halo.  The tree starts at $z=0$ and works backwards in time, branching into 
progenitor halos, until it reaches a starting redshift.  The halo merger rates 
used by the merger tree algorithm are those derived by \citet{lacey93}.  
The merger statistics that underpin the SA model are of particular
importance to $\PNM$, since the number of galaxies in any given halo should be
closely related to the formation and merger history of its progenitor halos.
Once a merger tree is created, a suite of analytic prescriptions is 
used to model the formation and evolution of galaxies in each progenitor halo, 
starting with the highest redshift progenitors and moving forward in time all 
the way to the single halo at $z=0$.

Each halo is given an NFW \citep{navarro96} dark matter density profile (with no
scatter in halo concentration) and an 
angular momentum drawn from a log-normal distribution.  Diffuse gas is assumed 
to be shock heated to the halo virial temperature during the formation process 
and to settle initially into a spherical distribution.  Gas that is dense enough 
to radiate its thermal energy before the halo experiences a major merger is 
assumed to accrete onto a centrifugally supported disk at the halo center.  This 
mechanism proceeds from the center of the halo outwards, since the cooling 
timescale is an increasing function of radius within the halo.  Cold gas that 
has settled onto the disk begins forming stars at a rate proportional to the 
total mass of cold disk gas and inversely proportional to an empirical 
timescale, which is described below.  The effects of stellar winds are modeled 
by returning a fraction of stellar mass into the cold gas phase, and the effects 
of feedback are parameterized by reheating a fraction of the cold gas to the 
halo virial temperature and ejecting it from the disk.  The SA model keeps track 
of how much gas is in the hot, cold, and stellar phases at any given time and 
traces chemical enrichment of the gas by following the exchange of metals among 
these three phases.  

If at any point during this process the merger tree contains a merger between 
two halos, the most massive galaxy is assumed to become
the ``central'' galaxy of the merged halo and 
any other galaxies present become ``satellites''.  Each of these satellite 
galaxies is assigned a random orbit and a timescale on which dynamical 
friction causes it to merge with the central galaxy.  If such a galaxy merger 
happens, the central galaxy's evolution may be mildly or severely disrupted, 
depending on the mass ratio $M_{\mathrm{sat}}/M_{\mathrm{cen}}$ of the merging 
galaxies.  If $M_{\mathrm{sat}}/M_{\mathrm{cen}} \geq 0.5$, then the merger is 
classified as ``major,'' and the two galaxies form an elliptical galaxy,
with their remaining gas consumed in a 
single burst of star formation.  If 
$0.5 > M_{\mathrm{sat}}/M_{\mathrm{cen}} \geq 0.25$, then the star formation 
burst still happens, but the central galaxy disk is not destroyed.  
Finally, if
$M_{\mathrm{sat}}/M_{\mathrm{cen}} < 0.25$, then the satellite's gas and stars 
are added to that of the central galaxy without disrupting it.
There are several adjustable parameters in the SA model, and their values
are chosen so that the model reproduces some observed properties of the
local galaxy population, in particular the galaxy luminosity function.
Other observables then serve to test the model.  
The SA parameters are {\it not} adjusted on the basis of galaxy clustering
measurements. 
\cite{yoshida02} and \cite{helly03} compare the galaxy
properties predicted by SA calculations and SPH simulations, and in
these tests they adjust the SA parameters to mimic the physical assumptions
and numerical resolution of the simulations.
Here we have chosen to take both methods ``as is''; the SA model 
incorporates its standard set of physical processes, and its parameters
are adjusted on the basis of observations.

The SA model computes many observable properties of galaxies such as
luminosities, sizes, colors, metallicities, and morphological types. 
The properties that interest us here are, in addition to halo membership, 
the total baryonic mass and the mass-weighted mean stellar age.  We will 
also make use of the SA bulge-to-disk ratios in \S\ref{MassAge}.  While 
the nature of the semi-analytic model used in this work is identical to 
that described by \citet{cole00}, the parameters differ from those of
their fiducial model because we have adopted different cosmological
parameters.  In particular, we adopt the baryon density parameter
$\Omega_b=0.02h^{-2}=0.0473$ used in the SPH simulation, where \cite{cole00} 
used $\Omega_b=0.02$, and we are forced to alter other model parameters to 
maintain a good match to local galaxy luminosity functions.  The star 
formation timescale in the semi-analytic model is now described by
\begin{equation}
\tau_\star = \tau^0_\star (V_{\rm disk}/200\kms)^{\alpha_\star},
\end{equation}
with $\tau^0_\star=3$Gyr and $\alpha_\star=-2.5$, and the feedback parameter 
$V_{\rm hot}$ is increased to $250\kms$.  We also impose a minimum star 
formation timescale of 1 or 25 Myrs for quiescent and bursting star formation 
respectively.  We adopt the Salpeter IMF and a recycled fraction $R=0.373$
(our choice of IMF is not particularly important for this work since
we only consider the total stellar mass of galaxies, rather than their
photometric properties).  The core radius of the gas density profile in the 
model halos is initially set to $2/3$ of the NFW scale radius (twice as 
large as in \citealt{cole00}).  We inhibit cooling of gas in dark matter 
halos with virial velocities below $60\kms$ to mimic the effects of an 
ionizing background.  Finally, the critical mass ratios of merging galaxies
that determine when elliptical galaxies and bursts of star formation are 
produced have changed to the values described above (from 
$M_{\mathrm{sat}}/M_{\mathrm{cen}} > 0.3$ in \citealt{cole00}).
The dependence of SA model predictions on input parameters and modeling
assumptions has been examined extensively in other papers (e.g.,
\citealt{kauffmann93}; \citealt{cole94}, 2000; \citealt{somerville99};
\citealt{benson02}).  We will not attempt such an investigation here, but 
we note that the good agreement we find between the HOD predictions of the 
SA model and the SPH simulations, two radically different calculational 
methods, suggests that the SA predictions themselves will not be sensitive
to the modeling details, at least for models that are tuned to reproduce the 
observed galaxy luminosity function.

As in the SPH case, we construct three galaxy samples of different space 
densities by only including galaxies above a baryonic mass threshold.  In the 
SA model these thresholds are $\Mbmin=$ 
$1.45\times10^{10} M_{\odot}$, $3.14\times10^{10} M_{\odot}$, and
$5.67\times10^{10} M_{\odot}$ for galaxy samples with space densities of
$\ng=0.02\hvol$, $\ng=0.01\hvol$, and  $\ng=0.005\hvol$, respectively.  
These mass thresholds are approximately a factor of four lower 
than the corresponding SPH thresholds, a difference that we discuss
in the following section.

\subsection{Galaxy Mass and Correlation Functions} \label{models:xi}

\begin{figure}
\plotone{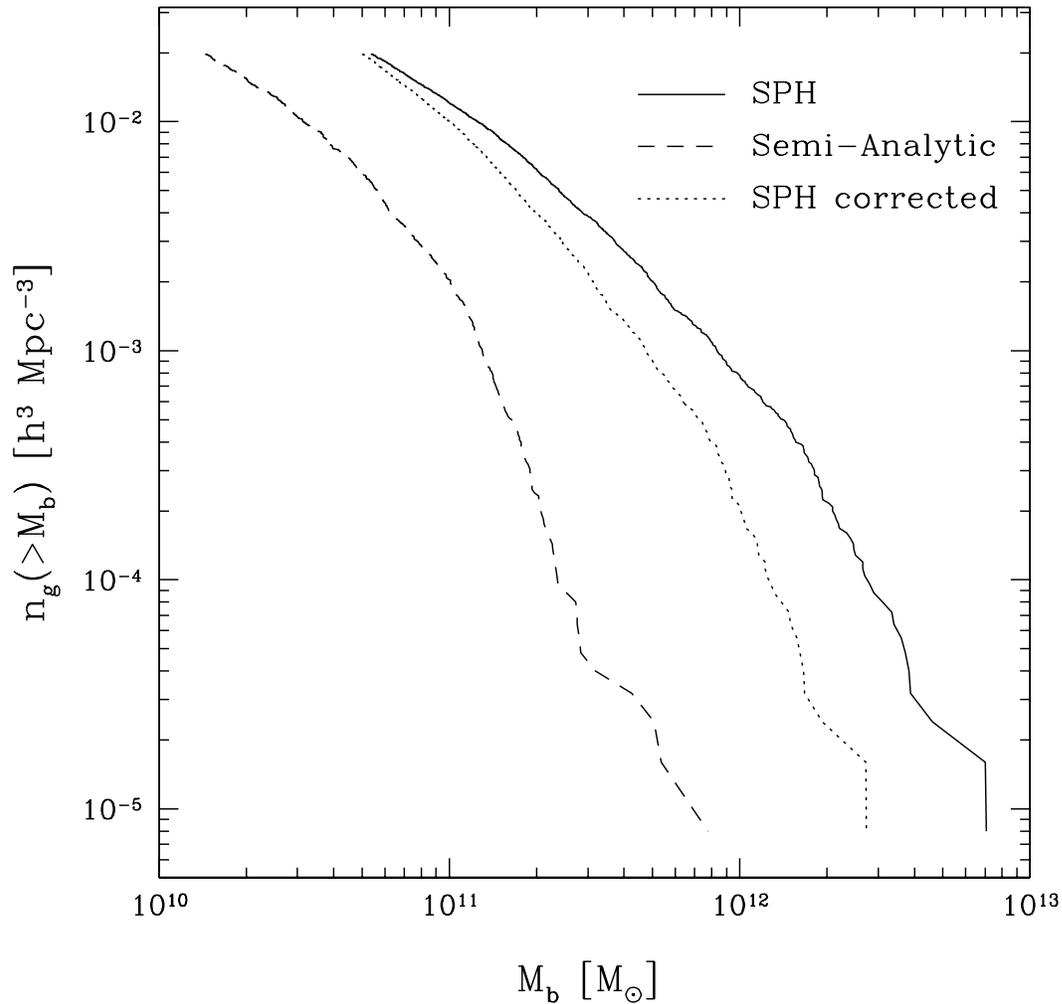}
\caption{Galaxy mass functions for the SPH and Semi-Analytic models.  
Solid and dashed curves
represent cumulative baryonic mass functions of SPH 
and SA galaxies, respectively.  The dotted curve incorporates a correction
for finite resolution effects on the SPH galaxy mass function,
estimated by comparing two simulations of a $22.222\hmpc$ cube.
} 
\label{fig:ngM}
\end{figure}

Figure~\ref{fig:ngM} shows cumulative baryonic mass functions for SPH and SA 
galaxies, starting at the thresholds that define our main analysis samples,
with space density $\ng=0.02\hvol$.  
At any given space density, the SA galaxies are less massive 
by a factor of $4-10$.  The parameters of the SA model, primarily
those controlling the gas core radius and stellar feedback, are
chosen to produce a good fit to the observed galaxy luminosity
function.  As we will show later, the baryonic mass of the SA galaxies
hosted by a given dark matter halo rarely exceeds 25\% of the total
mass of baryons within the halo virial radius.
In the SPH simulations, on the other hand, low mass halos frequently
host a galaxy with $M_b \sim (\Omega_b/\Omega_m)M_h$, corresponding
to 100\% of the baryon mass within the virial radius.
For observationally motivated choices of the stellar mass-to-light
ratio, the resulting Tully-Fisher (\citeyear{tully77}) relation
is in reasonable accord with observations, but the predicted
luminosity function is too high (Katz et al., in preparation).

The growing gap between the SA and SPH mass functions in 
Figure~\ref{fig:ngM} is partly a consequence of numerical resolution
effects in the SPH simulation.  A galaxy-by-galaxy comparison of two
simulations in a $22.222\hmpc$ cube, the $2\times 128^3$ particle
simulation mentioned in \S\ref{models:sph} and a $2\times 64^3$ 
simulation that has the same resolution as our $2\times 144^3$,
$50\hmpc$ run, indicates that the lower resolution calculations,
which also do not include the effects of a photoionizing UV background,
yield approximately correct masses for galaxies near the $64m_{\rm SPH}$
threshold but systematically overestimate the masses of larger galaxies,
probably because of the 2-phase interface effects discussed by
\citet{pearce01}, \citet{croft01}, and \citet{springel02}.
The dotted curve in Figure~\ref{fig:ngM} shows the result of correcting
the $50\hmpc$ cube mass function for this effect, using an empirical
formula derived from the $22.222\hmpc$ simulations (Fardal et al., in
preparation).  This curve represents our best guess at the mass
function we would obtain with a $2\times 288^3$ particle simulation
of a $50\hmpc$ cube including a UV background field.
With this rescaling, the gap between the SPH
and SA mass functions is a roughly constant factor of $3-4$. 
As shown by \cite{weinberg02a}, the weak lensing mass-to-light ratios
obtained from the $50\hmpc$ simulation (with this rescaling)
agree fairly well with those inferred by \cite{mckay01} from SDSS data.
The conflicting implications of the weak lensing and luminosity function
comparisons remain a puzzle, at least if our choice of cosmological
parameters is correct.

\begin{figure}
\plotone{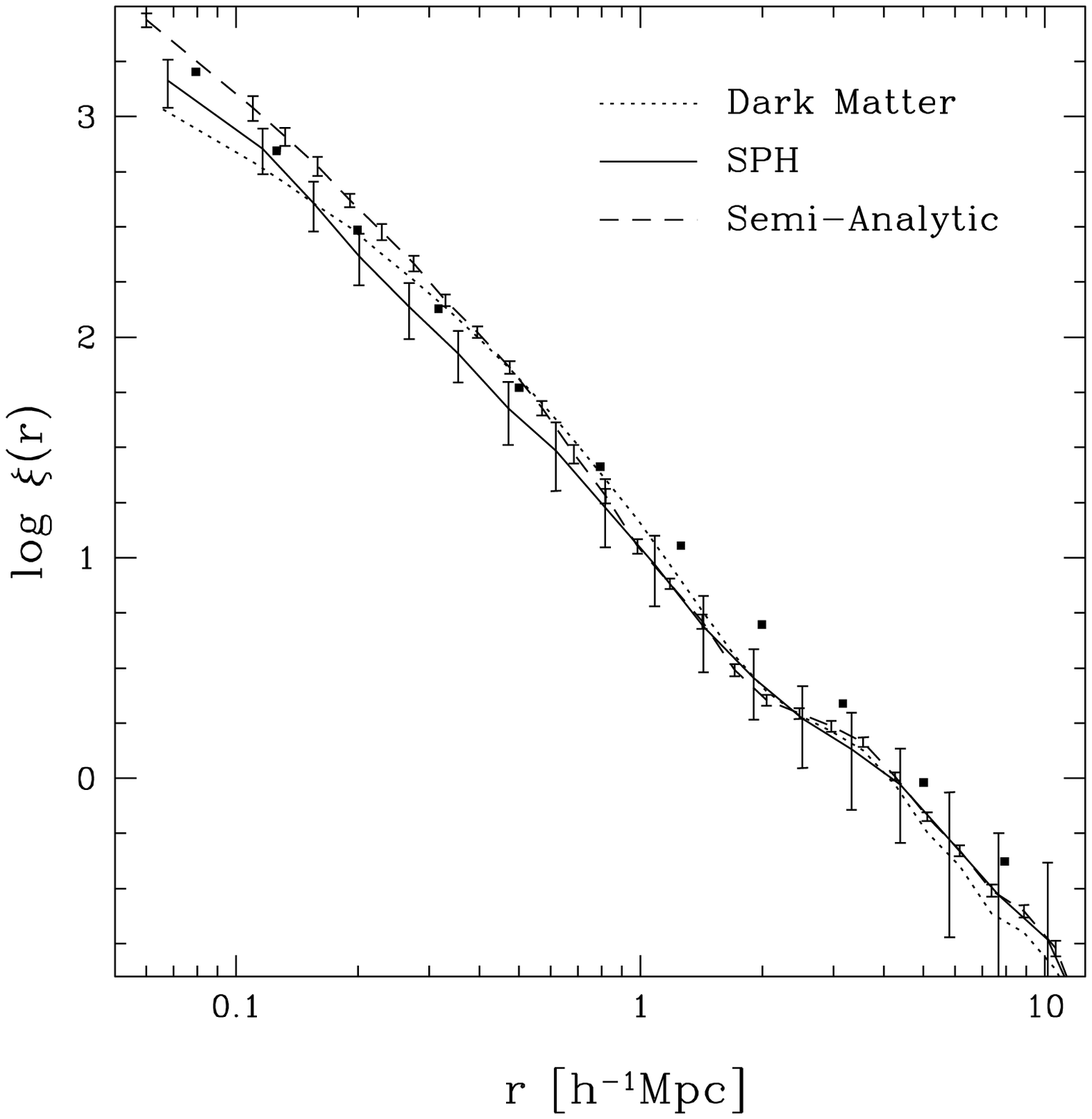}
\caption{Two-point correlation functions of dark matter and model galaxies.
The curves show the correlation function of the dark matter in the SPH
simulation (dotted), the SPH galaxies (solid), and the SA galaxies (dashed).
The points show the fitted power-law relation for 2dF $L_*$ galaxies from 
\citet{norberg02}: $\xi(r)=(r/4.9)^{-1.79}$.
The SPH and SA galaxy samples have a space density of $\ng=0.02\hvol$ and
thus correspond to a population of galaxies less luminous than $L_*$.
The error bars 
shown for the SPH correlation function are the errors in the mean 
estimated from jackknife resampling using the eight octants of the cube,
and they thus include an estimate of ``cosmic variance'' of the finite
number of coherent structures in the simulation volume.
The SA correlation function shown is the average over 10 realizations of
the SA model and the error bars are the uncertainty in the mean.  The SA
error bars thus represent the uncertainty in $\PNM$ predicted by the model
but do not include cosmic variance.
} 
\label{fig:xi}
\end{figure}

We select galaxy samples above baryon mass thresholds, but we characterize
these samples by their space density $\ng$ rather than the mass threshold
itself.  The membership in a given sample, and thus the HOD, would 
be unchanged by any monotonic rescaling of galaxy masses.  The discrepancy
of SA and SPH baryon mass functions reflects the combined impact of 
differing physical assumptions (e.g., regarding stellar feedback),
the approximations in the SA method, and the numerical limitations of
the SPH simulation.  We will not attempt to disentangle these contributions
here, but we will show that the two approaches nonetheless give similar
predictions for the clustering of galaxy samples at common space density.

Figure~\ref{fig:xi} shows the most commonly studied galaxy clustering
statistic, the two-point correlation function, for the two $\ng=0.02\hvol$
galaxy samples and for the dark matter in the SPH simulation.  
We compute the correlation function of SA galaxies by populating the
SPH dark matter halos with SA galaxies assuming that, in every halo that 
contains one or more galaxies, the first galaxy is located at the center of 
mass of the halo and any remaining galaxies trace the dark matter distribution 
within the halo.  The plotted correlation function of SA galaxies is the mean 
$\xig$ of the 10 SA realizations, and the plotted error bars show the 
uncertainty in the mean;  the $1\sigma$ dispersion from one $\PNM$ realization
to another is a factor of $10^{1/2}$ larger.  The SA error bars therefore 
illustrate the uncertainty due to fluctuations in $\PNM$, but they do not
include uncertainty due to the finite number of large scale structures
in the $50\hmpc$ cube.  The SPH error bars, on the other hand, are computed
by jackknife resampling using the eight octants of the simulation cube 
(see \citealt{weinberg02a}), and they are dominated by the ``cosmic variance''
in these large scale structures.  For comparison, we show a power law with the 
parameters derived for $L_*$ galaxies in the 2dF redshift survey 
\citep{norberg02}.

The dark matter correlation function shows the steepening at $r\sim1\hmpc$
and levelling off at $r\lesssim0.3\hmpc$ that has been found in most N-body
studies (e.g., \citealt{jenkins98}, though the deviation from a power law 
there is stronger than here due to a larger value of $\sigma_8$).
Both the SPH and SA models have correlation functions that are closer to a
straight power law from $r\sim0.05-2\hmpc$, although, like the dark matter, 
they show a kink at that larger scale.  The difference between the model
galaxy and dark matter correlation functions is difficult to see because of the
small size of the simulation box.  This difference stands out more strongly
when we use the SPH and SA HODs to populate a larger volume N-body simulation
(shown in Fig.~\ref{fig:xiJen}).  The low amplitude of the model $\xig$ at 
scales larger than $\sim1\hmpc$ compared to the 2dF points in 
Figure~\ref{fig:xi} is partly a consequence of the particular realization of 
structure in this $50\hmpc$ volume \citep{weinberg02a} and partly due to the 
fact that our $\ng=0.02$ threshold corresponds to a population of galaxies less 
luminous than $L_*$.  The comparison of the SPH $\xig$ to observations is 
discussed at greater length by \citet{weinberg02a}.  The correlation functions 
of the two models differ slightly, with the SA model having a higher amplitude 
of $\xig$ on scales smaller than $\sim0.8\hmpc$ and a somewhat more pronounced 
feature around $\sim3\hmpc$.  From $\xig$ alone, it is difficult to say what 
differences in galaxy formation physics are responsible for these differences 
in clustering.  With the HOD analysis that follows, we will see that the small 
scale difference arises mainly from the greater representation of galaxies in 
the highest mass halos predicted by the SA model.


\section{Halo Occupation Probabilities $\PNM$} \label{pnm}

We now turn to the primary results of this study, comparison of
the $\PNM$ predicted by the SPH and SA models.  
Unless we specify otherwise, the 
results we show are for the galaxy samples constructed to have a space density 
of $\ng=0.02\hvol$, corresponding to galaxies brighter than $\sim 0.2L_*$ 
for the \citet{blanton01} $r$-band luminosity function or the 
\citet{norberg02b} $b_J$-band luminosity function.
  
\begin{figure}
\plotone{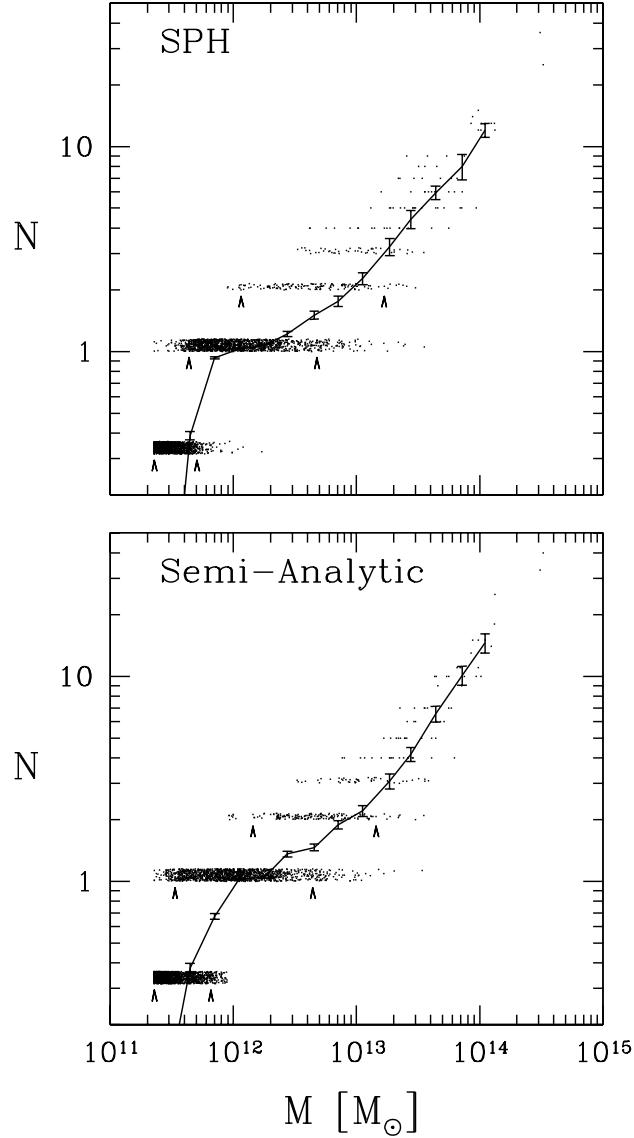}
\caption{Predicted $N(M)$ from the SPH and SA models.  Each point 
represents the number of galaxies above a baryonic mass threshold 
(selected to yield a galaxy population of space density $\ng=0.02\hvol$) 
that occupy a single dark matter halo in the SPH simulation (top panel) 
and a single realization of the SA model (bottom panel).  Points for 
halos that contain no galaxies are arbitrarily placed at log$N=-0.5$.  
A small random vertical scatter is added to points with $N<4$ 
to reduce saturation, and caret marks indicate the central $90\%$
of points with $N<3$.  The solid curves show the mean $\N$ and its 
uncertainty, computed in bins of log$M$.
} 
\label{fig:pnm}
\end{figure}

Figure~\ref{fig:pnm} shows $N(M)$ predicted by the SPH model (top panel) and
a single realization of the SA model (bottom panel).  Each point 
represents the number of galaxies in a specific halo, and the solid curve shows
the mean $\N$ and its statistical uncertainty, computed in bins of halo mass.  
The general features that can be seen in $\N_M$ (we use this notation to denote 
the mean occupation as a function of halo mass) for both models are a sharp 
drop in the fraction of halos that contain a galaxy for halos of mass less than 
$\sim 5\times 10^{11}\Msun$ and a slope that increases from roughly
$\N \propto M^{0.2}$ to $\N \propto M^{0.8}$ as $M$ gets larger.  This behavior 
is similar to that of the broken power-law $\N_M$ shown by BW 
to produce a good match to the observed galaxy correlation function 
(their Fig.~9).
For halos with $M \la 2\times 10^{12}\Msun$, the distribution $\PNN$
is close to a ``nearest integer'' (hereafter Nint) distribution with 
only two possible multiplicities, 0 and 1 or 1 and 2.\footnote{
In \cite{benson00} and BW, this is referred to 
as the ``Average'' distribution, since it is as close as one can 
come to $N=\N_M$ given that the former is integer valued and the
latter is not.  The general definition is
$p(N_l|\N)=1-(\N-N_l),$ $p(N_l+1|\N) = \N-N_l$,
where $N_l$ is the integer satisfying $N_l \leq \N < N_l+1$,
with $p(N|\N)=0$ for all other values of $N$.}
The scatter in $\PNN$ is larger at higher masses ---
for example, a $10^{13}\Msun$ halo may have $1-4$ galaxies --- but
we will show that the scatter remains significantly below that of
a Poisson distribution up to $M\sim 2\times 10^{13}\Msun$.
Overall, the results in Figure~\ref{fig:pnm} are in 
qualitative agreement with results from other hydrodynamic simulations 
(\citealt{white01}; \citealt{yoshikawa01}; \citealt{pearce01}) and 
semi-analytic calculations (\citealt{kauffmann99}; \citealt{benson00}; 
\citealt{somerville01}).

\subsection{Mean Halo Occupation} \label{Navg}

\begin{figure}
\plotone{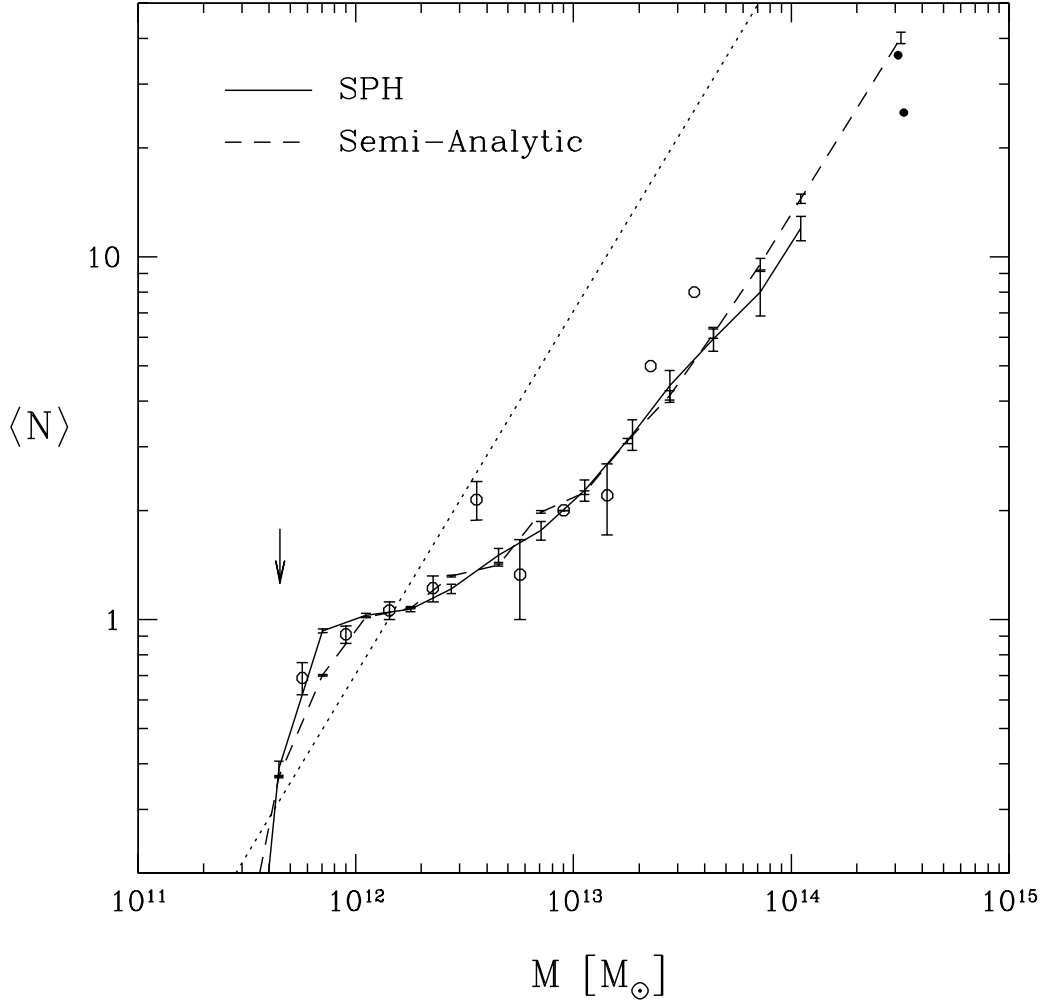}
\caption{Predicted $\N_M$ relation from the SPH and SA models.  The curves 
show the mean $\N$ and its uncertainty, computed in bins of log$M$, for the 
SPH (solid) and SA (dashed) models.  Both models have minimum galaxy 
baryonic masses selected to yield a galaxy population of space density 
$\ng=0.02\hvol$.  In the SA case, we use 10 random realizations of each 
halo, resulting in smaller error bars for $\N_M$.  In the SPH case, we show 
the two most massive halos as points.  For purpose of comparison, we show 
the relation $\N \propto M$ (dotted line), normalized so that $N=1$ at the 
same mass as the SPH and SA models.  The arrow marks the halo mass that would 
contain a total baryonic mass equal to the minimum SPH galaxy baryonic mass, 
assuming the universal baryon to mass ratio.  Open circles show the $\N_M$
results obtained from a smaller volume simulation with a factor of eight 
higher mass resolution.  Error bars for these circles are only shown for
mass bins that contain at least three halos.
} 
\label{fig:N}
\end{figure}

We compare the $\N_M$ relations predicted by the SPH and SA models in 
Figure~\ref{fig:N}.  The result for the SA model is the average over the 10
SA realizations, resulting in smaller error bars, especially at 
$M\sim 3\times 10^{14}\Msun$ where there are only two halos in the SPH
simulation.  Also shown is the relation
$\N_M \propto M$, normalized so that $N=1$ at the same mass as the SPH and SA 
models (dotted line).  The agreement between the two predicted $\N_M$ relations 
is strikingly good.
They have roughly the same cutoff at low mass and the same shape 
and amplitude across more than two orders of magnitude in halo mass.  The only 
notable differences are that the SPH simulation predicts 
a slightly sharper cutoff at low masses and slightly fewer galaxies on average 
in the highest mass halos, though with a single SPH realization it is not
clear that this latter discrepancy is statistically significant.
Our matching of number densities requires that the integrals
$\ng = \int_0^\infty \N_M n(M) dM$, where $n(M)$ is the halo mass function,
be equal in the two models, but it clearly does not enforce the detailed
agreement seen in Figure~\ref{fig:N}.

We expect a low mass cutoff in $\N_M$ simply because lower mass halos do not
contain enough gas to form a galaxy of baryonic mass greater than $\Mbmin$,
the baryonic mass threshold of our sample.  
If the ratio of galaxy baryonic mass to halo mass were equal to the 
universal baryon-to-matter ratio, the minimum halo mass would be
$\Mmin=(\Omegam/\Omegab)\Mbmin$.  
The vertical arrow in Figure~\ref{fig:N} marks this halo mass for the value of 
$\Mbmin$ that defines the SPH galaxy sample.
Since this arrow nearly coincides with the sharp drop in $\N_M$,
we conclude that many low mass halos in the SPH simulation cool essentially
all of the gas within the virial radius into a single 
galaxy.\footnote{We are using the phrase ``within the halo virial radius''
somewhat loosely here.  In the SPH simulation, accretion along filaments
means that the Lagrangian (initial) volume of gas within the final virial
radius is larger and more irregular than that of the dark matter.
Some gas travels long distances along a filament to the central galaxy,
while some gas within the halo virial radius remains hot 
(see \citealt{katz94,katz02}).}  The $\N_M$ relation of the SA model also 
cuts off at this halo mass, but in this case the value of $\Mbmin$ is four 
times smaller, implying that these galaxies contain only $\sim 25\%$ of the 
baryons within the virial radius.  As discussed in \S\ref{models:xi}, we 
attribute this difference to the impact of the stellar feedback parameters 
in the SA model, which are chosen to match the observed galaxy luminosity 
function.  For the lowest mass halos in the SA model, feedback results in 
approximately four solar masses of gas ejected from galaxy disks for every 
one solar mass of stars formed.  Therefore, these halos cool gas with high 
efficiency, and feedback then determines how much cold gas is allowed to remain 
in galaxies and form stars.  The SA low mass cutoff is nearly as sharp as 
the SPH cutoff because the feedback mechanism is tightly correlated with 
halo mass, so galaxies in SA halos never retain more than $\sim 25\%$ of 
their gas.

\begin{figure}
\plotone{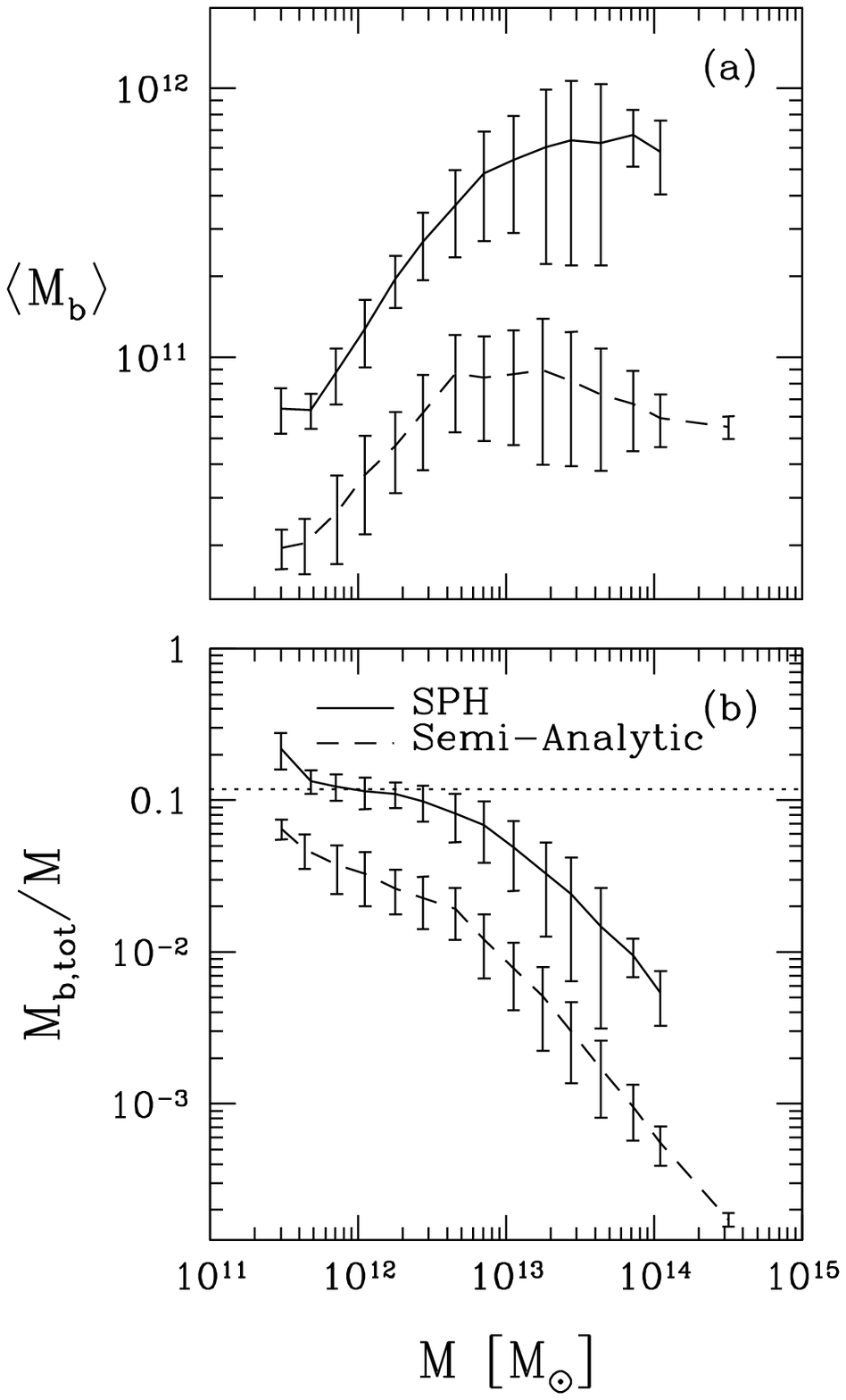}
\caption{(a) Mean baryonic galaxy mass per halo $\left<M_b\right>$, and
(b) fraction of total halo mass contained in galaxy baryons 
$M_{b,\mathrm{tot}}/M$.  The average of these quantities is computed in 
bins of log$M$ for the SPH (solid) and SA (dashed) models.  Also shown is 
the $1\sigma$ scatter in the relations.  The dotted line in panel (b) shows 
the universal baryon fraction $\Omegab/\Omegam$.  The relation between 
$\left<M_b\right>$ and $M_{b,\mathrm{tot}}$ in each halo is simply 
$M_{b,\mathrm{tot}}=N \left<M_b\right>$, where $N$ is the number of
galaxies in the halo. 
} 
\label{fig:Mgavg}
\end{figure}

Following the low mass cutoff, there is a low occupancy regime ($\N\lesssim2$) in
which the mean number of galaxies rises slowly, growing from one to two
over a decade in halo mass.  In this regime, halos ``spend'' their
larger gas supplies on building a more massive galaxy rather than
building multiple low mass galaxies.
Figure~\ref{fig:Mgavg}a demonstrates this point, showing
the average baryonic mass $\left<M_b\right>$ of galaxies in each halo,
averaged in bins of halo mass.  Although there is
a large offset in galaxy mass between the two models, both the SPH 
and SA models predict a steady increase in $\left<M_b\right>$ in the low
occupancy regime, leveling off at $M\sim 10^{12.5}-10^{13}\Msun$.  
Continuing to more massive halos, we enter a
high occupancy regime where $\N_M$ 
steepens, though the slope remains less than unity except perhaps at the very 
highest masses.  
This steepening marks a transition from a regime where much of the cooling
gas is channeled to one or two galaxies to a regime in which halos
are built by merging smaller halos whose pre-existing galaxies survive,
for the most part, as distinct entities.  The transition presumably
reflects the relative timescales for gas cooling and major halo mergers.
A mismatch between the scales of filaments and galaxies in high mass
halos may also play a role in this transition, with filaments of large 
geometrical cross-section no longer able to funnel cooled gas directly onto 
galaxies (see \citealt{katz02}).  Dynamical friction may also be more effective 
at bringing together the galaxies of merged halos in the low mass regime.
Figure~\ref{fig:Mgavg}b shows the fraction of total halo 
mass that is in the form of galaxy baryons, averaged in bins of halo mass.  
In the high halo mass regime, the efficiency of converting gas to 
galaxies drops with increasing halo mass, keeping the slope of $\N_M$
below unity.  Some of the gas may, of course, be going into galaxies
below our mass threshold, rather than not cooling at all.


We can encode these physical ideas in a fitting formula for $\N_M$,
\begin{equation}
\N_M = K \left(\frac{M}{\Mcrit}\right)^\alpha 
 \left[ 1+\left(\frac{M}{\Mcrit}\right)^\mu \right]^{\frac{\beta-\alpha}{\mu}}
 \left( 1 - \mathrm{exp}\left[-\left(\frac{M}{\Mmin}\right)^\nu\right] \right),
\label{eqn:Nfit}
\end{equation}
which has a cutoff, a low and high mass regime, and a transition between them.
$\Mmin$ sets the location of the low mass cutoff, $\nu$ determines how sharp it 
is, $\alpha$ and $\beta$ are the power-law slopes of $\N_M$ in the low and 
high mass regimes, $\Mcrit$ is the mass where the slope changes, and $\mu$ 
determines the speed of the transition.  The particular form of the cutoff is 
chosen because it gives a good match to the numerical results.  In place of 
the normalization constant $K$, one can specify the mass $M_1$ of halos that 
have a mean occupancy of one, $\N_M=1$. 
Table~1 lists the best fit values of these parameters for the SPH and SA 
models.  The last column in the table lists the maximum logarithmic error of the
fitting formula for log$\N$, $\Delta$(log$\N)_{\mathrm{max}}$, relative to the 
numerical results.  It is not surprising that we can fit the numerical data 
with a seven parameter function, and we could almost certainly find something 
with fewer parameters that would also work.  We choose this form because it 
seems a natural description of the results with some basis in the physical 
interpretation.  These fitting functions are useful for bootstrapping the 
predictions of these models onto larger volume N-body simulations, as we will 
do in \S\ref{xiJen} below.

One natural concern with the SPH predictions is the impact of finite
resolution.  In particular, one might worry that the effects of dynamical
friction are overestimated in low mass halos, and that an excessive
galaxy merger rate in this regime could be partly responsible for the
very shallow slope of $\N_M$ at low occupancy.
Open circles in Figure~\ref{fig:N} show the $\N_M$ results obtained
from the $22.222\hmpc$ simulation described in \S\ref{models:sph}.
Each halo now has eight times more dark matter particles than in
the $50\hmpc$ cube, and the baryonic mass threshold now corresponds
to the mass of 512 SPH particles rather than 64.  While the small
volume of the higher resolution simulation leads to poor representation
of high mass halos, the agreement in the low mass regime is excellent,
and the one $M\sim 3\times 10^{14}\Msun$ halo in the small box has
similar occupation to those in the large box.  
The agreement of numerical results across a factor of eight in
mass resolution, and the agreement between the SPH and SA results, 
suggests that the mean occupation function shown in Figure~\ref{fig:N}
is a secure prediction of the current theory of galaxy formation,
given our adopted cosmological parameters.

\subsection{Factorial Moments and $P(0|M)$} \label{PNNavg}

We now turn our attention to higher order moments of $\PNM$.  
These moments influence galaxy clustering on small scales, where the 
number of galaxy pairs, triples, and so forth within a single halo 
becomes important.  
For example, the two-point 
correlation function has a 1-halo term that depends on the second factorial
moment $\NN_M = \sum_{N=0}^\infty N(N-1)\PNM$.  The three-point 
correlation function has a 1-halo term that depends on $\NNN_M$ 
and a 2-halo term that depends on $\NN_M$. 
\cite{benson00} and BW show that a sub-Poisson $\PNN$ distribution
makes it much easier to produce a correlation function of the
observed power-law form; the alternative is to put galaxies into
halos of implausibly low mass (see also 
\citealt{seljak00,peacock00,scoccimarro01}).
The $\PNN$ distribution is therefore an important prediction of 
galaxy formation theories, in addition to $\N_M$.

\begin{figure}
\plotone{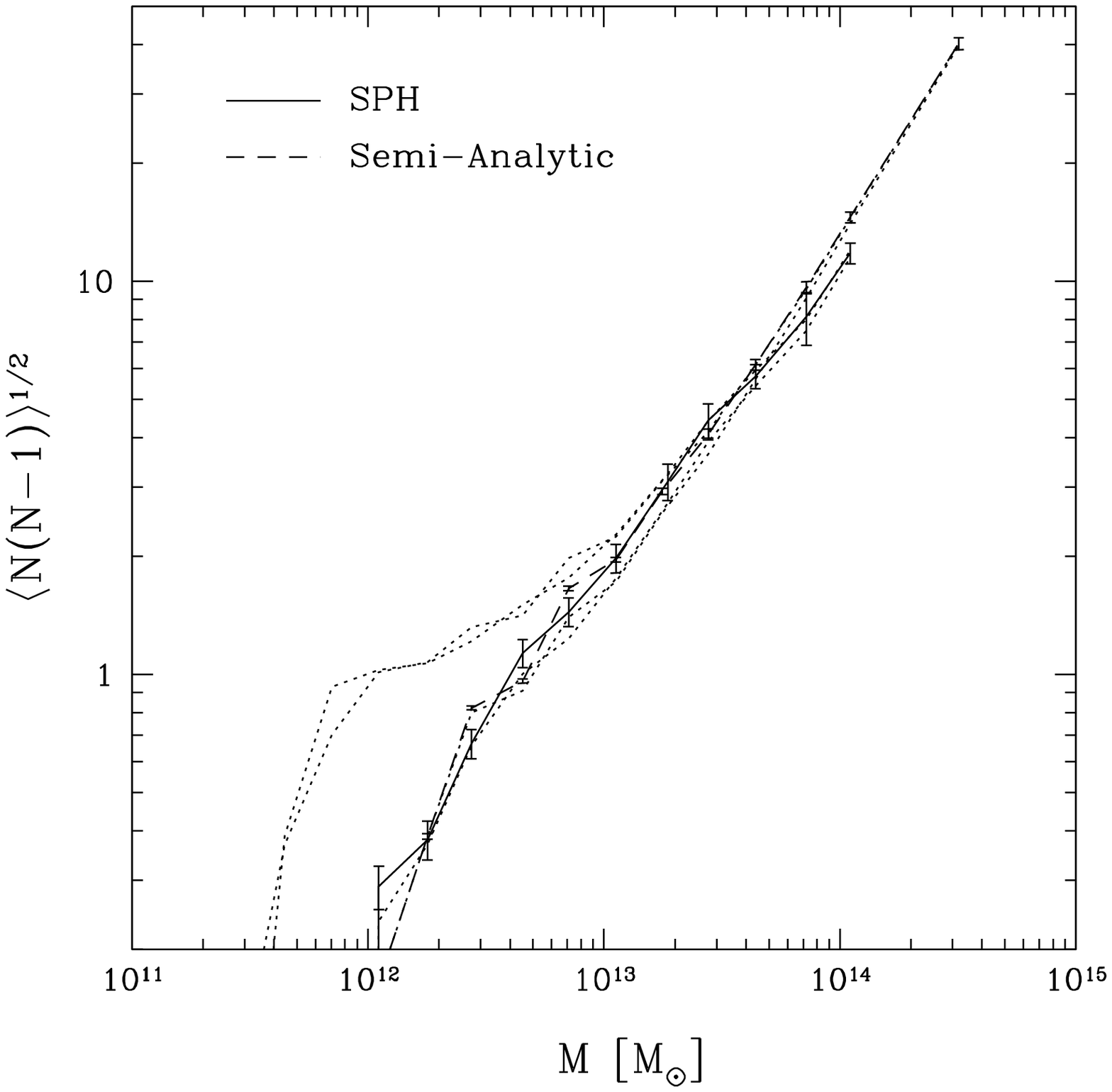}
\caption{Predicted $\NN_M^{1/2}$ relation from the SPH and SA models.  
The curves show the mean $\NN^{1/2}$ and its uncertainty, computed in 
bins of log$M$, for the SPH (solid) and SA (dashed) models.  Also shown 
for each model are the bracketing cases of Poisson and nearest-integer 
$\PNN$ (dotted lines above and below $\NN_M^{1/2}$, respectively), derived 
from equations~(\ref{eqn:xibar1}) and~(\ref{eqn:xibar2}).
} 
\label{fig:NN}
\end{figure}

The solid and dashed lines in Figure~\ref{fig:NN} show 
$\NN_M^{1/2}$, the square-root of the mean number of galaxy pairs in halos
of mass $M$, for the SPH and SA calculations, respectively.
As in Figure~\ref{fig:N}, the agreement of the two models is 
remarkably good.  The upper set of dotted curves shows the relation
$\NN^{1/2}=\N$, which would be expected for Poisson $\PNN$ distributions.
The lower set of dotted curves shows the corresponding prediction
for nearest-integer $\PNN$.  Clearly the distributions predicted by the
SPH and SA models are much narrower than Poisson distributions when the
occupation number is low, and they are close to the maximally narrow Nint
distributions.  In particular, halos that on average contain $0-1$ 
galaxies almost never contain 2, and halos that on average contain
$1-2$ galaxies rarely contain 3 (see Fig.~\ref{fig:pnm}).
Since the cutoff in $\N_M$ at $\N<1$ is quite sharp, the former
result can be understood largely in terms of mass supply: halos
with $\N<1$ do not have enough cold baryons to make two galaxies
above the mass threshold.  However, $5\times 10^{12}\Msun$ halos have
enough baryonic material to make $\sim 10$ galaxies above the SPH
baryon mass threshold, and they rarely make even three or four.
The sub-Poisson width of $\PNN$ at these scales suggests that halos
in this mass regime have a relatively narrow range of formation and
accretion histories, a degree of regularity that the SPH and SA
calculations evidently agree upon.

We can quantify the effects of sub-Poisson fluctuations on halo pair
counts via the quantity $\omega = (\Nsqr-\N^2)/\N$, which is equal to 
one for a Poisson distribution, less than one for narrower distributions,
and greater than one for broader distributions.  In terms of this parameter,
the second factorial moment is $\NN = \N^2+\N(\omega-1)$.  
A nearest-integer distribution has $\omega \approx 1$ for $\N\ll 1$
falling to $\omega=0$ at $\N=1$.  The Nint value of $\omega$ is 
exactly zero at all higher integer values of $\N$, and it rises
slightly above zero for non-integer values 
\citep{yang03}.\footnote{BW incorrectly implied that the Nint value of
$\omega$ is zero for all $\N$.  A $\delta-$function $\PNN$ has
$\omega \equiv 0$, but since a $\delta$-function
is not restricted to integer values 
of $N$, it is not an acceptable model for this physical situation.}
From the mean pair count results plotted in Figure~\ref{fig:NN},
we find that the SPH and SA models predict an $\omega$ that falls
quickly from one to zero in the cutoff regime with $\N<1$, then
rises steadily from zero to one as halo masses increase from 
$10^{12}\Msun$ to $10^{14}\Msun$, with an approximately linear
trend of $\omega$ with log$M$.  Thus, we find a steady trend 
from Nint pair counts in the low occupancy regime to Poisson
pair counts in the high occupancy regime.  Note, however, that
the fractional difference between Nint and Poisson pair counts
is large at low $\N$ but small at high $\N$.
One can see from Figure~\ref{fig:NN} that the Nint model is never
very far, in a logarithmic sense, from the predicted pair counts,
while the Poisson model is much too high for $\N \la 1.5$.
Although we do not show them here, the pair counts predicted by
the high resolution SPH simulation are in good agreement with those
predicted by the large volume simulation.

\begin{figure}
\plotone{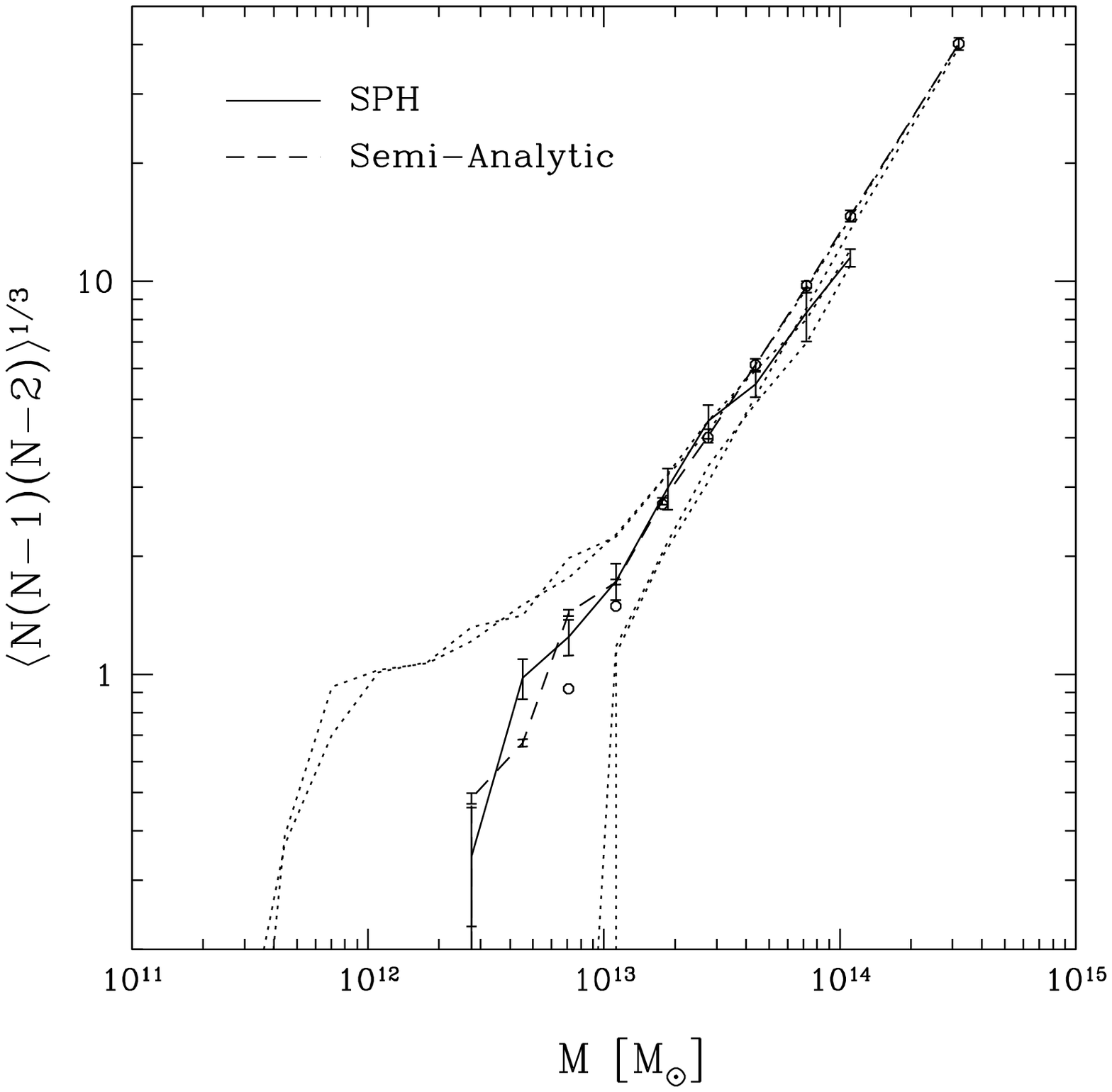}
\caption{Predicted $\NNN_M^{1/3}$ relation from the SPH and SA models.  The 
curves show the mean $\NNN^{1/3}$ and its uncertainties, computed in bins 
of log$M$, for the SPH (solid) and SA (dashed) models.  Also shown for 
each model are the bracketing cases of Poisson and nearest-integer $\PNN$ 
(dotted lines above and below $\NNN_M^{1/3}$, respectively), derived from 
equations~(\ref{eqn:xibar1}) and~(\ref{eqn:xibar2}).  The open points show 
$\N(1+3\bar{\xi}_2)^{1/3}$ for the SA model, where $\bar{\xi}_2$
is the second-order volume averaged connected correlation as defined in 
\S~\ref{PNNavg}.
} 
\label{fig:NNN}
\end{figure}

Figure~\ref{fig:NNN} shows $\NNN^{1/3}$, the cube-root of the mean
number of galaxy triples in halos of mass $M$.  The SPH and SA
predictions, shown by the solid and dashed lines, respectively,
again agree very well.  A Poisson $\PNN$ has 
$\NNN^{1/3}=\N$, shown by the upper dotted curves in Figure~\ref{fig:NNN}.
The predicted triple counts are substantially sub-Poisson, though in this
case they are higher than those for a nearest-integer distribution shown
by the lower dotted curves.  It is interesting to examine this behavior
in terms analogous to those used in studies of counts-in-cells statistics
(e.g., \citealt{colombi00}).\footnote{We thank S.~Colombi and J.~Fry for
suggesting this analysis.}
We define volume-averaged, connected correlations $\xibar_2(M)$
and $\xibar_3(M)$ by the relations
\begin{equation}
\NN  =  \N^2(1+\xibar_2), \qquad
\NNN =  \N^3(1+3\xibar_2+\xibar_3).
\label{eqn:xibar1}
\end{equation}
A Poisson $\PNN$ has $\xibar_2=\xibar_3=0$.  A Nint distribution with
$N_l \leq \N < N_l+1$ has
\begin{equation}
\xibar_2 = -\frac{N_l(N_l+1)}{\N^2} + \frac{2N_l}{\N} -1, \qquad
\xibar_3 = -\frac{2N_l(N_l^2-1)}{\N^3} + \frac{6N_l^2}{\N^2} - \frac{6N_l}{\N} +2,
\label{eqn:xibar2}
\end{equation}
which reduce to the $\delta-$function values 
$\xibar_2=-1/\N$ and $\xibar_3=2/\N^2$ for integer values of $\N$.
We can obtain values of $\xibar_2(M)$ from the pair counts plotted
in Figure~\ref{fig:NN}.  Open circles in Figure~\ref{fig:NNN} show
$\N(1+3\xibar_2)^{1/3}$ for the SA model, i.e., the triple counts
predicted if we incorporate the measured sub-Poisson behavior of
the pair counts but set $\xibar_3=0$.  Since these points match the
actual triple counts better than either the Poisson prediction or
the full Nint prediction with negative $\xibar_3$, we conclude that 
the sub-Poisson statistics of galaxy assembly in the SA and SPH 
models have a direct impact on pair counts in halos but influence
triple counts mainly via this indirect impact on pair counts.
We have not investigated higher order factorial moments, but the
results here suggest the conjecture $\xibar_N\approx 0$ for $N>2$.

\begin{figure}
\plotone{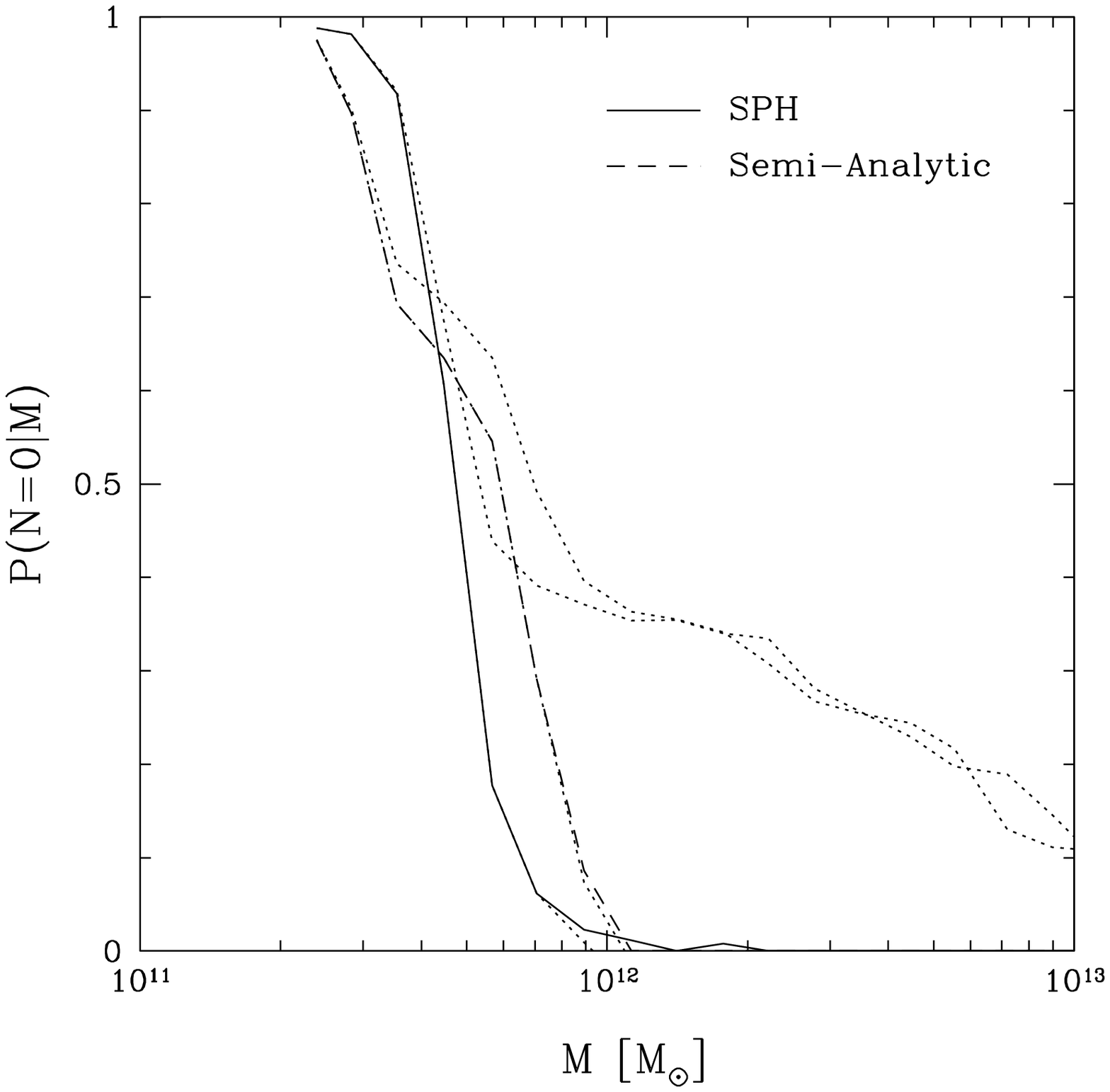}
\caption{Predicted $P(0|M)$ distribution from the SPH and SA models.  The curves
show the probability $P(0|M)$ that a halo of mass $M$ contains zero galaxies
above the baryonic mass threshold, computed in bins of log$M$.  
Top two dotted curves show the expected distributions for a Poisson $\PNN$, 
which predicts $P(0|M)=\mathrm{exp}(-\N_M)$.  The model results are much 
better described by the nearest-integer prediction $P(0|M)={\rm max}(1-\N_M,0)$ 
(dotted curves that deviate only at the very tail ends of $P(0|M)$).
} 
\label{fig:P0}
\end{figure}

In low mass density regions of the universe, the halo mass function 
is shifted to lower masses (e.g., \citealt{mo96}), making it less likely that 
these regions contain massive halos.  The question of whether these regions are
void of galaxies thus depends on the probability $P(0|M)$ that 
a halo of mass $M$ contains zero galaxies of a specified class; the void 
probability itself is high if $P(0|M)$ is high for all halos below the 
exponential cutoff scale of the shifted mass function (see BW, \S 4.3).
Therefore, for calculating galaxy void statistics, the HOD property
of greatest importance is $P(0|M)$.  Solid and dashed curves in 
Figure~\ref{fig:P0} show $P(0|M)$ for the SPH and SA models, respectively.  
The halos with $N=0$ are in general not empty of galaxies, but the galaxies
that they contain are below our baryonic mass threshold $\Mbmin$.
Both $P(0|M)$ curves drop from one to zero over the mass range
$2\times10^{11}$ --- $10^{12}\Msun$, corresponding to the rise in $\N_M$
from zero to one in Figure~\ref{fig:N}.
The top two dotted curves show the expectations for Poisson statistics,
$P(0|M)=\exp(-\N_M)$.  Analogous to the factorial moment results,
it is clear that the probability of a high mass halo being empty is far
lower than Poisson statistics would imply.
In fact, the SPH and SA results are almost perfectly described by 
the Nint distribution (bottom two dotted curves), for which 
$P(0|M)={\rm max}(1-\N_M,0)$; only the low amplitude ($P\la 0.02$) 
tail of the SPH prediction for 
$8\times 10^{11}\Msun \la M \la 2\times 10^{12}\Msun$ deviates
noticeably from this result.
The sharpness of $P(0|M)$ is further testimony to the regularity
of the galaxy formation process in low mass halos.  In the SPH simulation,
halos with $M \la 3.5\times 10^{11}\Msun$ almost never contain a
galaxy above the baryonic mass threshold, while halos with 
$M\ga 7\times 10^{11}\Msun$ almost always do.
The transition is slightly more gradual in the SA model, as one might
expect, since the SPH cutoff is driven by the universal baryon fraction
while the SA cutoff is determined by the physical processes that
suppress gas cooling, whose operation depends to some extent on the halo's
formation history.
The sub-Poisson nature of $P(0|M)$ slightly decreases the probability
of finding large, empty voids (see BW, Fig. 14c).  However, for galaxy samples
defined by luminosity rather than baryonic mass, variations in 
stellar populations may significantly soften the transition from
empty halos to occupied halos.

\subsection{Mass and Age Dependence of $\PNM$} \label{MassAge}

\begin{figure}
\plotone{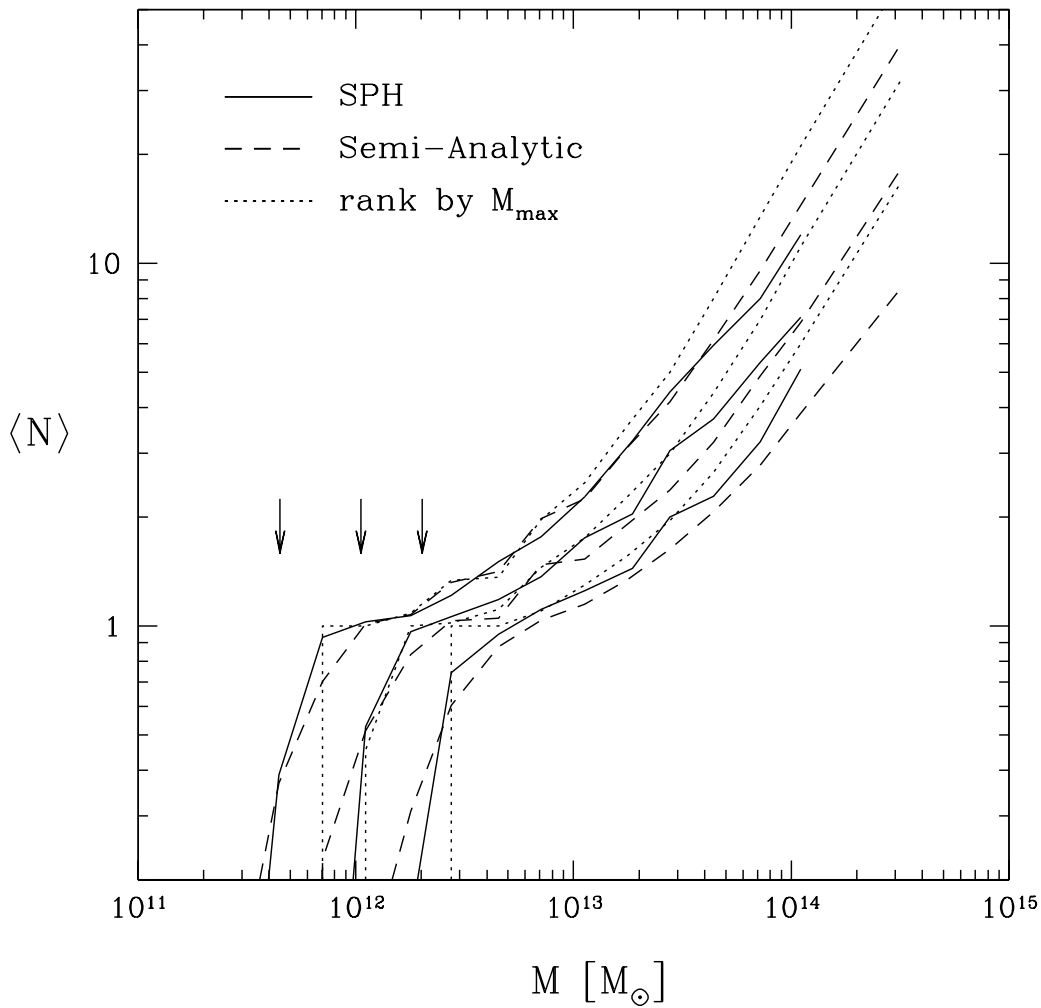}
\caption{Predicted $\N_M$ from the SPH and SA models, as a function of galaxy 
space density.  Both models have minimum galaxy baryonic masses selected to 
yield galaxy populations of space densities $\ng=0.02\hvol$ (top curves), 
$0.01\hvol$ (middle curves), and $0.005\hvol$ (bottom curves).  The three 
vertical arrows mark the halo masses that would contain a total baryonic 
mass equal to the minimum SPH galaxy baryonic mass, under the assumption of 
a universal baryon to mass ratio.  Dotted curves show $\N_M$ for the SA model
when galaxies are selected above a threshold in $\Mmax$ (defined in 
\S~\ref{conclusions}), rather than their baryonic mass.
} 
\label{fig:Nn}
\end{figure}

Figure~\ref{fig:Nn} shows $\N_M$ predicted by the SPH and SA models for 
samples with three different baryon mass thresholds $\Mbmin$, with 
corresponding space densities $\ng=0.02$, 0.01, and $0.005\hvol$.  
Luminosity-thresholded samples with the same space densities would have 
$L_{\rm min}\approx 0.2L_*$, $0.45L_*$, and $0.75L_*$, assuming the 
$r$-band luminosity function of \cite{blanton01} or the $b_J$-band 
luminosity function of \cite{norberg02b}.  Not surprisingly, the minimum 
halo masses are higher for higher baryon mass thresholds, and high mass 
halos necessarily contain fewer galaxies above these higher thresholds.
The low mass cutoffs for the SPH $\N_M$ continue to be dictated
by the universal baryon fraction, as indicated by the vertical 
arrows in Figure~\ref{fig:Nn}.  The SA cutoffs are always somewhat
softer than the SPH cutoffs, but they occur at nearly the same
halo masses, which is not surprising as it would otherwise be difficult
for the two models to have the same galaxy density $\ng$.

\begin{figure}
\plotone{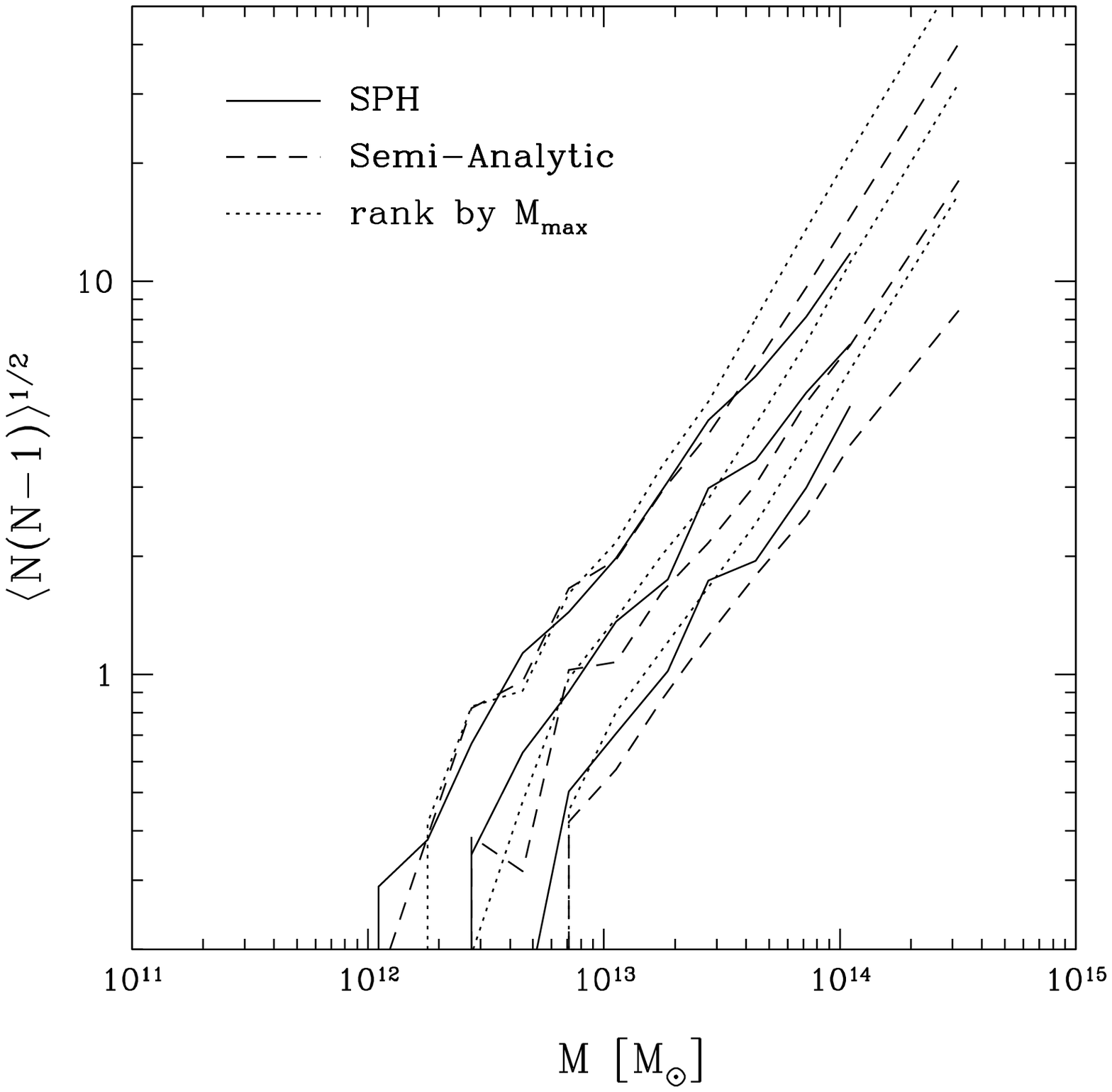}
\caption{Predicted $\NN_M^{1/2}$ from the SPH and SA models, as a function of 
galaxy space density.  Curves as in Fig.~\ref{fig:Nn}.
} 
\label{fig:NNn}
\end{figure}

The striking aspects of Figure~\ref{fig:Nn} are the good agreement
between the SPH and SA predictions at all three space densities
and the extent to which the change in $\N_M$ is well described by 
a simple horizontal shift along the log$M$ axis.  In terms of
the fitting function~(\ref{eqn:Nfit}), the effect of an increased
baryonic mass threshold is, approximately, to multiply the
mass scales $\Mmin$, $M_1$, and $\Mcrit$ by the ratio $f_M$ of
the new and old values of $\Mbmin$, while the slopes $\alpha$ and $\beta$
remain roughly the same.  Table~1 lists the values of $\Mbmin$ and the 
fitting function parameters for the different space density samples, and 
it also lists the values of $\N_M$ at $M=\Mcrit$, demonstrating that 
the transition from the low occupancy regime to the high occupancy 
regime occurs at $\N\sim 1.4-2$ in all cases.  Figure~\ref{fig:NNn} 
shows that the good agreement of SPH and SA calculations and the 
horizontal shift nature of the dependence on mass threshold also 
applies to the second factorial moment $\NN$.  We have computed 
$\omega$ for the various samples and find that the transition from Nint 
to Poisson pair counts always happens at roughly the same mean
occupancy.  The value of $\omega$ typically rises from 0 to 0.5 as 
$\N$ goes from $\sim1$ to $\sim2$, and $\omega \approx 1$ for 
$\N \gtrsim 5$.  We have examined lower baryonic mass thresholds in the 
high resolution, $22.222\hmpc$ SPH simulation (not shown)
and find analogous horizontal shifts of $\N_M$ and $\NN_M$.

Dotted curves in Figures~\ref{fig:Nn} and~\ref{fig:NNn} represent
results of a calculation using halo merger tree properties instead
of baryonic mass to select SA galaxy samples.
We will discuss this calculation in \S\ref{conclusions}.


\begin{figure}
\plotone{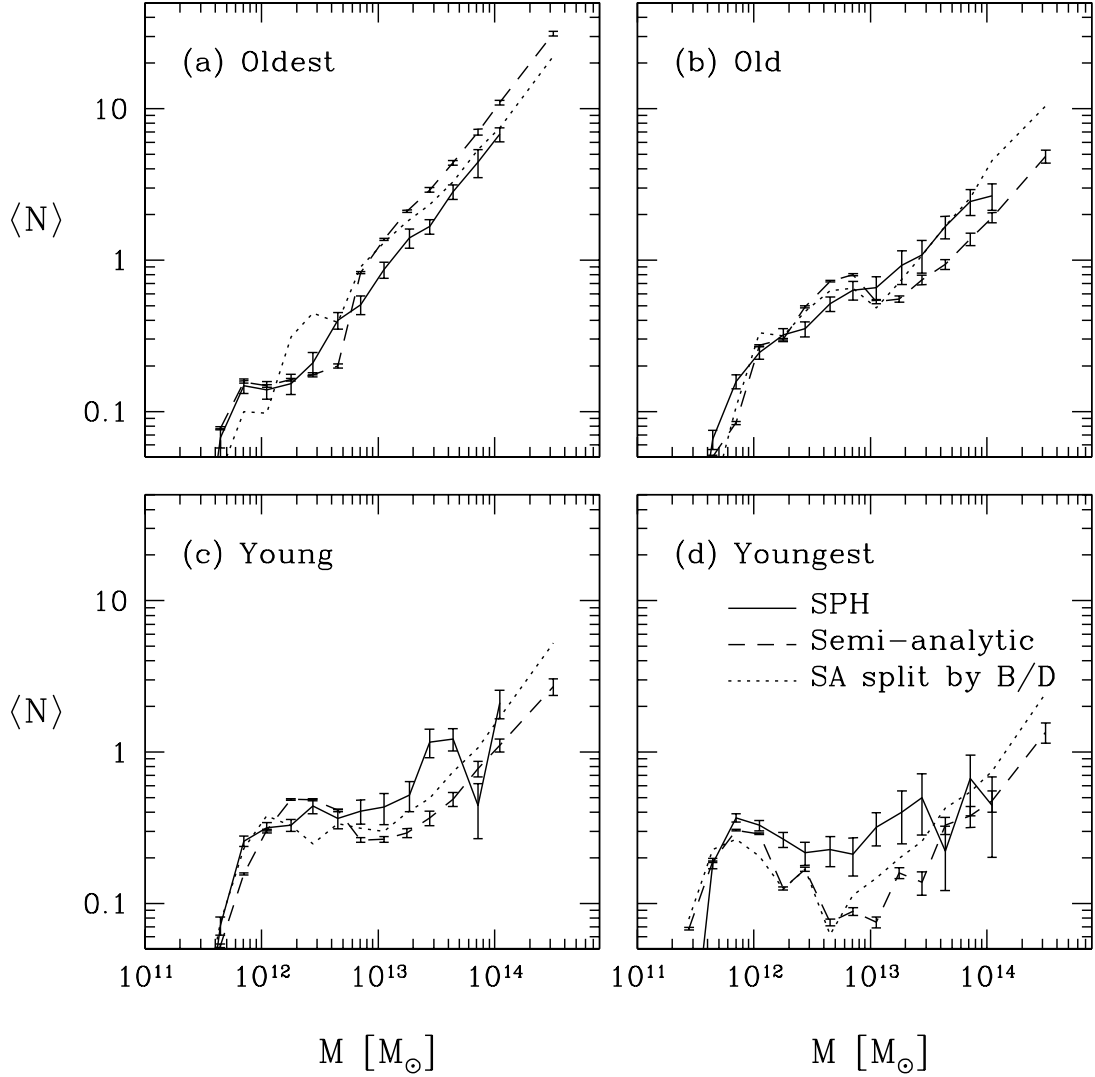}
\caption{Dependence of $\N_M$ on galaxy age.  The four panels correspond
to four age quartiles for the $\ng=0.02\hvol$ samples, from oldest (a)
to youngest (d).  In each panel, solid and dashed lines show results
for SPH and SA galaxies, respectively, with error bars showing the
error on the mean in each mass bin.  Dotted curves show SA galaxies
classified into quartiles based on bulge-to-disk ratio, from bulge
dominated (a) to disk dominated (d).
}
\label{fig:Nage}
\end{figure}

The clustering of galaxies is well known to depend on galaxy color or 
spectral type and on galaxy morphology (see \citealt{norberg02,zehavi02}, 
and numerous references therein).  Reproducing the observed type dependence 
of clustering is an important test for theories of galaxy formation.  
Figure~\ref{fig:Nage} shows the mean occupation $\N_M$ predicted by the SPH 
and SA models for galaxies in four quartiles of stellar population age, 
which should correlate tightly with color or spectral type.  The SPH ages 
are median mass-weighted stellar ages, while the SA ages are mean 
mass-weighted stellar ages, because these quantities are straightforward to 
compute in the two analysis codes.  Although there are differences between 
the SPH and SA predictions in individual age quartiles, the agreement on the 
qualitative dependence of $\N_M$ on galaxy age is remarkably good.  Old 
galaxies have a steep $\N_M$ relation, with most galaxies occupying high 
multiplicity halos, whereas young galaxies have a shallow $\N_M$ and reside 
primarily in single galaxy halos.  In the language more commonly used to 
describe the environmental dependence of galaxy types, both calculations 
predict that old/red/early-type galaxies reside preferentially in clusters 
and that young/blue/late-type galaxies reside mainly in the field.  This 
result emerges from two aspects of the galaxy formation physics.  First, 
gravitational collapse and galaxy assembly begin earlier in the overdense 
regions that eventually form massive halos.  Second, gas accretion largely 
shuts off when a galaxy's parent halo merges into a more massive halo, 
starving the galaxy of the fuel that it would need to make young stars.  
The galaxies that are in high mass halos today started forming their stars 
early, and they stopped forming them some time ago.

The SPH simulation does not resolve galaxy morphology, but the SA
model does track morphology, assigning post-merger stellar populations
to bulge components as described in \S\ref{models:sa}.  Galaxy color and
morphological type are correlated with environment in similar ways.
Nonetheless, the physics that determines morphology is different from
the physics that governs stellar population age, and it is interesting
to ask whether the SA model predicts a morphological dependence of
clustering that is distinguishable from the age dependence.  Dotted curves 
in Figure~\ref{fig:Nage} show the SA predictions for galaxy quartiles defined 
by bulge-to-disk ratios (in stellar mass) rather than by stellar 
population age.  While there is a fair amount of galaxy-by-galaxy scatter 
between mean stellar age and bulge-to-disk ratio, the two ways of 
characterizing galaxy type display similar mean occupation functions.  The 
SA model thus predicts that the morphological dependence of clustering will 
closely track the age dependence of clustering.

\begin{figure}
\plotone{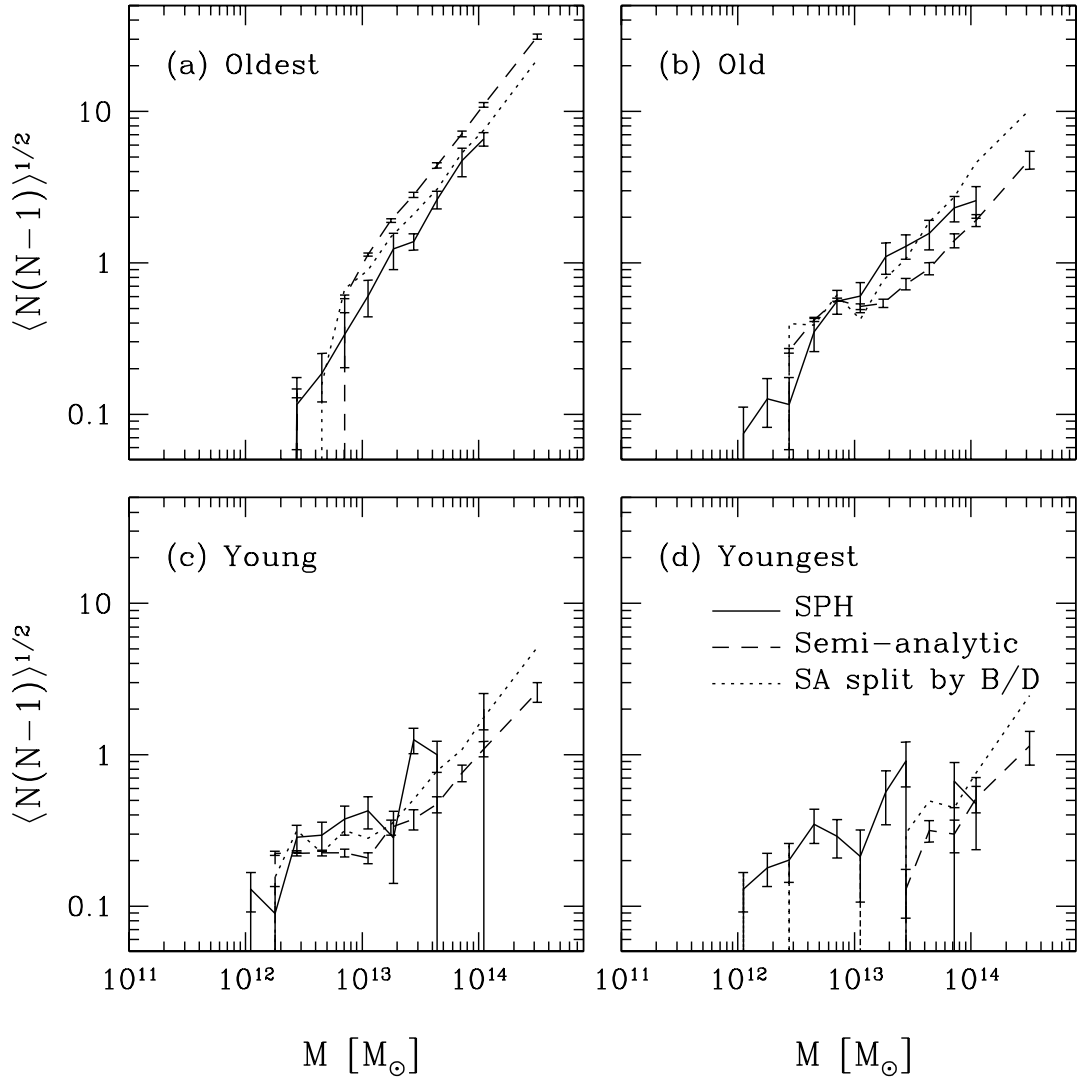}
\caption{Dependence of $\NN_M^{1/2}$ on galaxy age and (for the SA model)
bulge-to-disk ratio.  Panels and curves as in Fig.~\ref{fig:Nage}.
}
\label{fig:NNage}
\end{figure}
\begin{figure}
\plotone{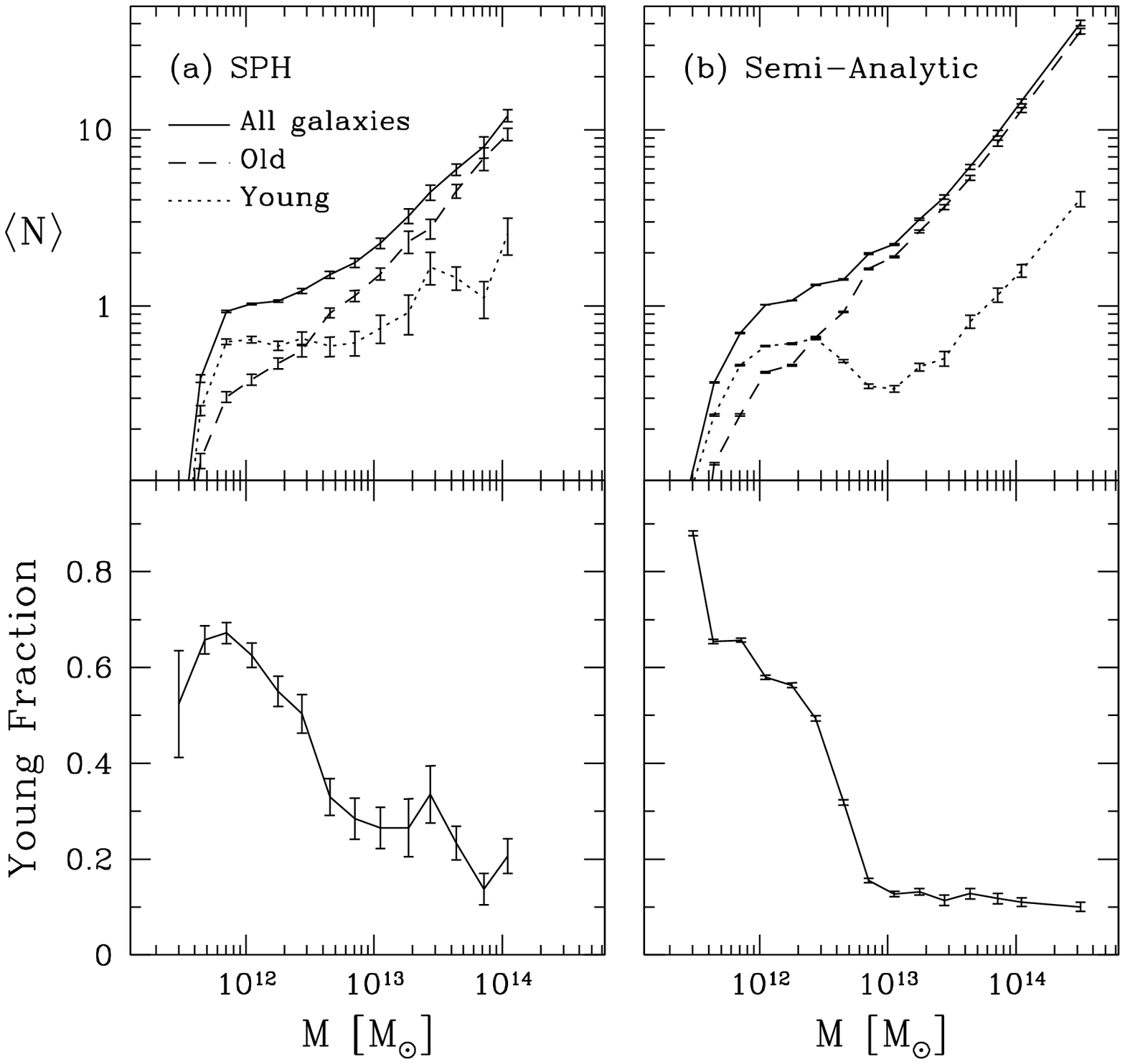}
\caption{Fraction of young galaxies as a function of halo mass.  Top panels 
show the mean halo occupation of old (dashed curves), young (dotted curves),
and all (solid curves) galaxies in the (a) SPH and (b) SA models, when the
$\ng=0.02\hvol$ sample is divided into two equal halves according to age.  
Bottom panels show the mean fraction of young galaxies in halos as a function 
of halo mass.  Error bars show the uncertainty in the mean, calculated
in halo mass bins.
}
\label{fig:Nage2}
\end{figure}

Closer inspection of Figure~\ref{fig:Nage} reveals interesting structure in the
SA $\N_M$ curves that is significant relative to the error bars.  For the 
youngest quartile, $\N_M$ has a maximum at $M\approx 10^{12}\Msun$,
then falls before rising again at $M \ga 10^{13}\Msun$.  The next two
quartiles show similar peaks in $\N_M$, at $M\approx 3\times 10^{12}\Msun$
and $M\approx 8\times 10^{12}\Msun$.  These peaks represent the contribution
of the halos' central galaxies, the ages of which are correlated with the 
halo masses.  Young central galaxies form in low mass halos, and the mean 
occupation for these galaxies declines toward higher halo mass as the central 
galaxy age moves out of the youngest quartile and into the second quartile, 
and so on for the third.  The oldest quartile has a very low plateau for low 
halo masses because the galaxies of these halos are almost always younger;
then there is a sharp rise when the halo mass gets high enough that 
central galaxies can be in the oldest quartile.  A similar peak at low halo 
masses for young/blue galaxies is also seen in the GIF semi-analytic models 
of \cite{kauffmann99}, and it is modeled as a Gaussian bump by \cite{sheth01} 
and \cite{scranton02}.  While the behavior of the SPH $\N_M$ curves is similar, 
the structure is much less pronounced, indicating that the SPH calculation 
produces more scatter between halo mass and central galaxy age.  Some of this 
difference could arise from the different age definitions, though there is no 
obvious reason that median ages would exhibit more scatter than mean ages.
The existence of local maxima in $\N_M$ means that our fitting 
formula~(\ref{eqn:Nfit}) does not describe the age quartile results accurately, 
so we do not attempt to perform fits for these samples.

Figure~\ref{fig:NNage} shows that the SPH and SA models predict qualitatively 
similar pair counts $\NN_M$ for the age quartiles, though the results are 
fairly noisy.  As with the mean occupation, age and bulge-to-disk divisions 
produce similar results for the SA model.  Overall, the strong age dependence 
of $\PNM$ implies that both SPH simulations and SA models predict a strong 
dependence of galaxy clustering on stellar population age.  The similarity of 
the predicted trends indicates that these dependences will be similar for the 
two methods, for all clustering statistics.  

Top panels of Figure~\ref{fig:Nage2} show the mean occupation for young, old, 
and all galaxies in the SPH and SA models, when the $\ng=0.02\hvol$ sample is
divided into two halves according to age.  While the quartile division 
demonstrates the steadiness of trends with age, this division in two yields more 
statistically robust predictions that can be tested against, for example, the 
red and blue halves of a volume-limited galaxy sample.  For both the SPH and SA 
calculations, the mean occupation of older galaxies is close to a power law 
truncated at $\Mmin$, and the slowly rising regime of the total $\N_M$ curve 
comes from adding this power law to the much flatter occupation curve of young 
galaxies.  The SA model predicts 
that in the high halo mass regime ($M \gtrsim 10^{13}\Msun$), the power-law 
slope of $\N_M$ for both old and young galaxies is equal to that of 
$\N_M$ for all galaxies.  This suggests that the fraction of young to old 
galaxies is constant in this regime.  This behavior is seen in the bottom panel 
of Figure~\ref{fig:Nage2}b, which shows the fraction of young galaxies in halos
as a function of halo mass.  The SA model predicts that the young fraction
drops steadily from 90\% to 10\% as halo mass increases from the minimum
cutoff mass to $\sim 10^{13}\Msun$, and then levels off at 10\% for greater 
masses.  The SPH model prediction for the young fraction (bottom panel of
Fig.~\ref{fig:Nage2}a) shows no clear evidence for this high mass plateau, 
but it is too noisy to tell for sure.  The mass dependence of the SA young 
fraction differs from that of the ``late-type fraction'' parameterization 
assumed by \cite{vandenbosch02}, which is forced to zero at high halo masses.

\subsection{(No) Environmental Dependence of $\PNM$} \label{env}

The claim that the HOD formalism offers a complete description of
galaxy bias hinges on a key assumption:
if we know, statistically, how galaxies occupy halos of given mass, then
we can predict all aspects of galaxy clustering given a halo population.
This assumption would break down if halos of the same mass in 
different large scale environments had systematically different galaxy
populations.  The excursion set model \citep{bond91} predicts that
the statistics of a halo's progenitor population and merger history
depend only on its mass, not on its environment.  The SA model
employed here is based on excursion set merger trees, so 
it necessarily incorporates no direct dependence of galaxy population
on halo environment.
The analysis of N-body simulations by \cite{lemson99} shows that halos
of the same mass in different environments have similar properties
and formation histories, providing substantial support for this
approach.  Nonetheless, it is desirable to revisit this central issue
with a simulation that includes the gas dynamics and dissipation 
that play key roles in galaxy formation.

\begin{figure}
\plotone{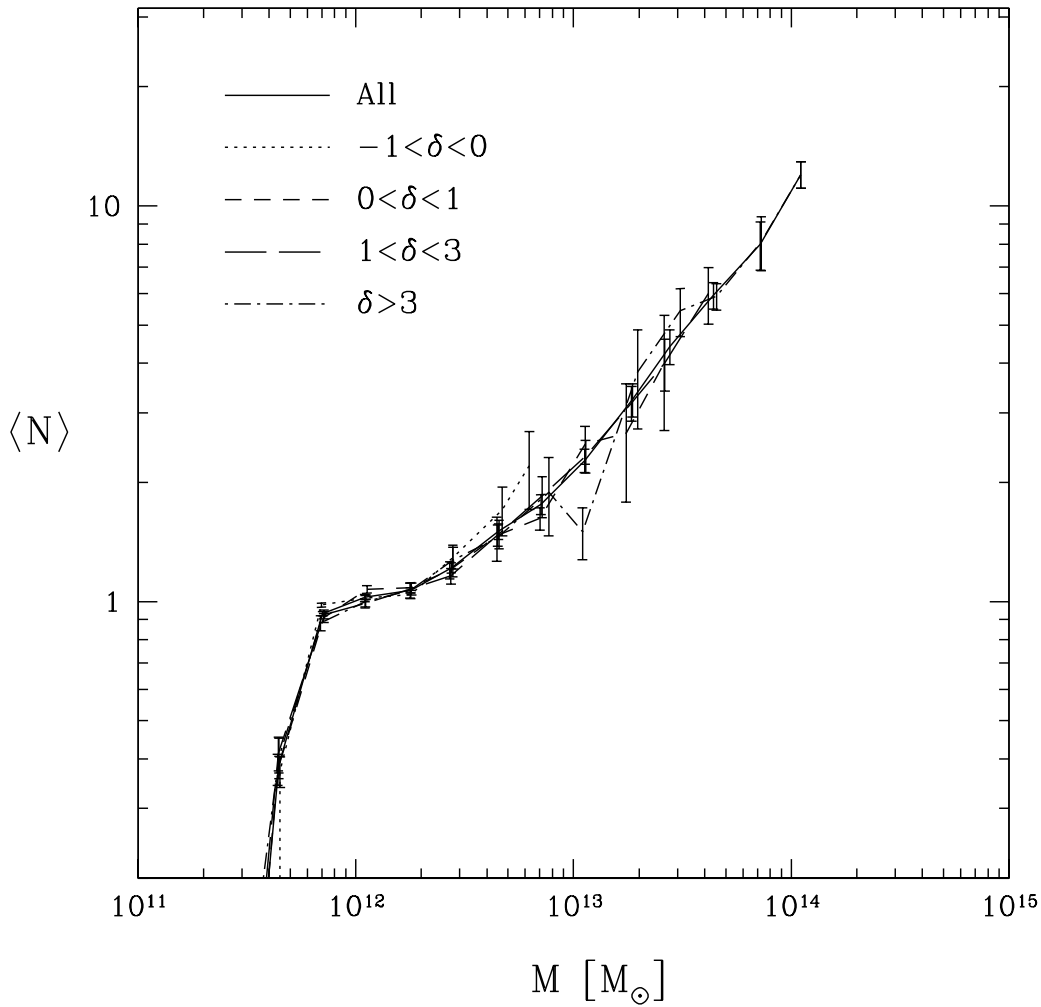}
\caption{Environmental dependence of $\N_M$ in the SPH model.  $\N_M$ and its
uncertainty is shown for all halos (solid curve) and for halos in four 
bins of dark matter density contrast $\delta$ as marked, where $\delta$ 
is computed in top-hat spheres of radius $4\hmpc$ around each halo.  
While the distribution of halo masses shifts with $\delta$, there is no 
discernible dependence of the galaxy occupation at fixed $M$
on the larger scale environment.
}
\label{fig:env}
\end{figure}

Figure~\ref{fig:env} shows $\N_M$ for the SPH galaxy populations of halos
in different bins of large scale density.
The density around each halo is found by smoothing the dark matter distribution
with a top-hat filter of radius $4\hmpc$.  As expected, the low 
density bins only probe $\N_M$ in the low halo mass regime, and successively 
higher density bins probe $\N_M$ to higher masses, demonstrating the expected
shift of the halo mass function with environment (see \citealt{mo96}).
The main result, however, is unambiguous: the mean halo occupation of 
galaxies is completely independent of the larger scale environment.  The 
same is true for $\NN_M$, although we do not plot that result here.  We thus 
conclude that, to the extent that the SPH simulation includes the most 
important macroscopic galaxy formation physics, the claim that the HOD 
is a complete formulation of galaxy bias is well founded.
The large scale relative bias between early and late type galaxies,
predicted by both the SA and SPH approaches \citep{benson00b,weinberg02a}
arises entirely from the dependence of the halo mass function on large
scale environment, not from changes in galaxy populations at fixed halo mass.


\section{The Distribution of SPH Galaxies Within Halos} \label{profile}

In this section we investigate the second component of the HOD: the relative
spatial and velocity distributions of galaxies and dark matter within halos.  
The spatial bias of galaxies within halos has a more limited impact on galaxy 
clustering than $\PNM$ because its effect is restricted to scales smaller 
than halo virial diameters.  Nevertheless, internal spatial biases can have 
an important effect on small scale clustering, such as the correlation
function at separations $r \la 0.3\hmpc$ (BW, Fig. 6).  Systematic differences 
between galaxy and dark matter velocity dispersions within halos, a.k.a. 
velocity bias, can have a major impact on redshift space clustering even at 
very large scales (BW, Figs. 15 and 16).  While the SA model treats presumed 
central galaxies differently from satellites, it does not directly predict 
spatial or velocity distributions within halos.  Clustering predictions based 
on the combination of SA modeling with N-body simulations often assume 
that the central galaxy resides at the halo center of mass and moves with 
the halo center of mass velocity, and that satellites trace the dark matter
spatial and velocity distribution (e.g., Kauffmann et al. 1997; 
Benson et al. 2000ab).  \citet{kauffmann99} adopt a somewhat different procedure,
in which satellite galaxies trace the most bound particle of their parent halo,
leaving some internal bias \citep{diaferio99}.  In the SPH simulation, we
can investigate internal biases in a calculation that includes the full effects
of gas dynamics, dynamical friction, and galaxy mergers.

\begin{figure}
\plotone{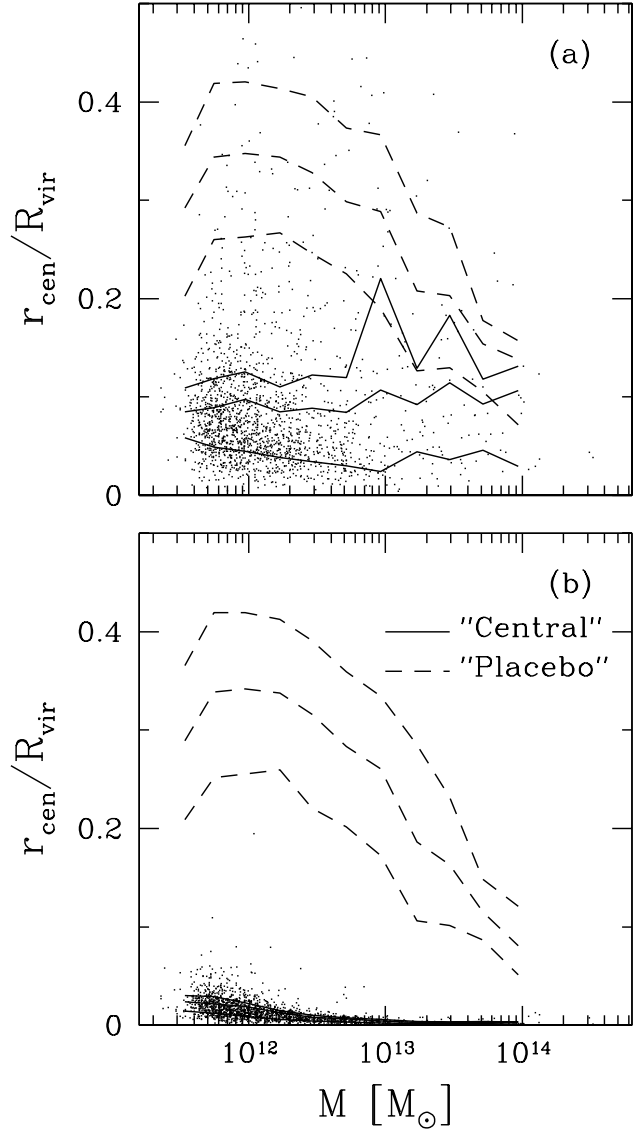}
\caption{(a) Distribution of ``central'' SPH galaxies' distances from their 
halo centers.  In each halo, the SPH galaxy closest to the halo's center of mass 
is tagged as the ``central'' galaxy.  Each point shows the distance of a
central galaxy from its host halo's center of mass, in units of the halo's
virial radius.  The middle solid curve shows the mean distance in bins of halo 
mass, and the outer two solid curves enclose 60\% of all SPH central galaxies.
The dashed curves show the same for dark matter ``placebo'' particles, which
represent the distribution of $\rcen/\Rvir$ that would be expected if 
SPH galaxies had the same distribution as dark matter within halos.
(b) Same as (a), but with the halo center identified
as the position of the most bound dark matter particle instead of the 
center of mass.
} 
\label{fig:Cen}
\end{figure}

Figure~\ref{fig:Cen}a plots the distribution $\rcen/\Rvir$ vs. $M$,
where $\Rvir$ is the halo virial radius and $\rcen$ is the distance
from the halo center of mass to the galaxy that is closest to it.
Each point represents an individual halo, the central solid curve shows the 
mean $\rcen/\Rvir$ in mass bins, and the upper and lower solid curves
enclose 60\% of the points.  On average, the centermost galaxy of
a halo resides within $0.1\Rvir$ of the halo center of mass.  
To obtain corresponding predictions for the case in which galaxies
trace mass within halos, we randomly select from each halo a number
of dark matter particles equal to the number of galaxies $N$.
We tag the selected particle closest to the center of mass as the
``placebo'' particle, and we repeat the process several times to
improve statistics.  Dashed curves in Figure~\ref{fig:Cen}a mark the
mean and central 60\% of the distributions of $\rpcb/\Rvir$ in bins
of halo mass.  Comparison to the solid curves demonstrates that the
center-most galaxies in the SPH halos are indeed a distinct, ``central''
population, whose proximity to the center of mass would not be expected
if galaxies randomly traced the dark matter.  

The distinction between central galaxies and placebo particles is less clear
at large $M$, but in these high mass halos, substructure near the virial radius
may shift the center of mass position.  The location of the most bound
dark matter particle or a local density maximum provides a more physically 
robust indicator of the halo center.  However, we began with the 
center-of-mass definition because the most bound dark matter particle
is almost guaranteed to lie near the center of a massive galaxy,
and with this definition of the halo center one might erroneously
infer the existence of ``central'' galaxies even if galaxies traced
the overall dark matter distribution within halos.  Figure~\ref{fig:Cen}b 
is the same as Figure~\ref{fig:Cen}a except that the halo center is 
identified as the position of the most bound dark matter particle
(specifically, the particle with the lowest potential energy, computed 
using the halo's dark matter particles only) instead of the center of mass.  
For the great majority of halos, this particle lies within $0.1\Rvir$ of 
the center of mass.  As expected, the $\rcen/\Rvir$ distribution is much 
narrower with the most-bound-particle definition of halo center, while the 
$\rpcb/\Rvir$ distribution is hardly changed.
Since Figure~\ref{fig:Cen}a convincingly establishes the existence of 
central galaxies, we will henceforth use the more reliable
most-bound-particle definition to identify which galaxies are central.

\begin{figure}
\plotone{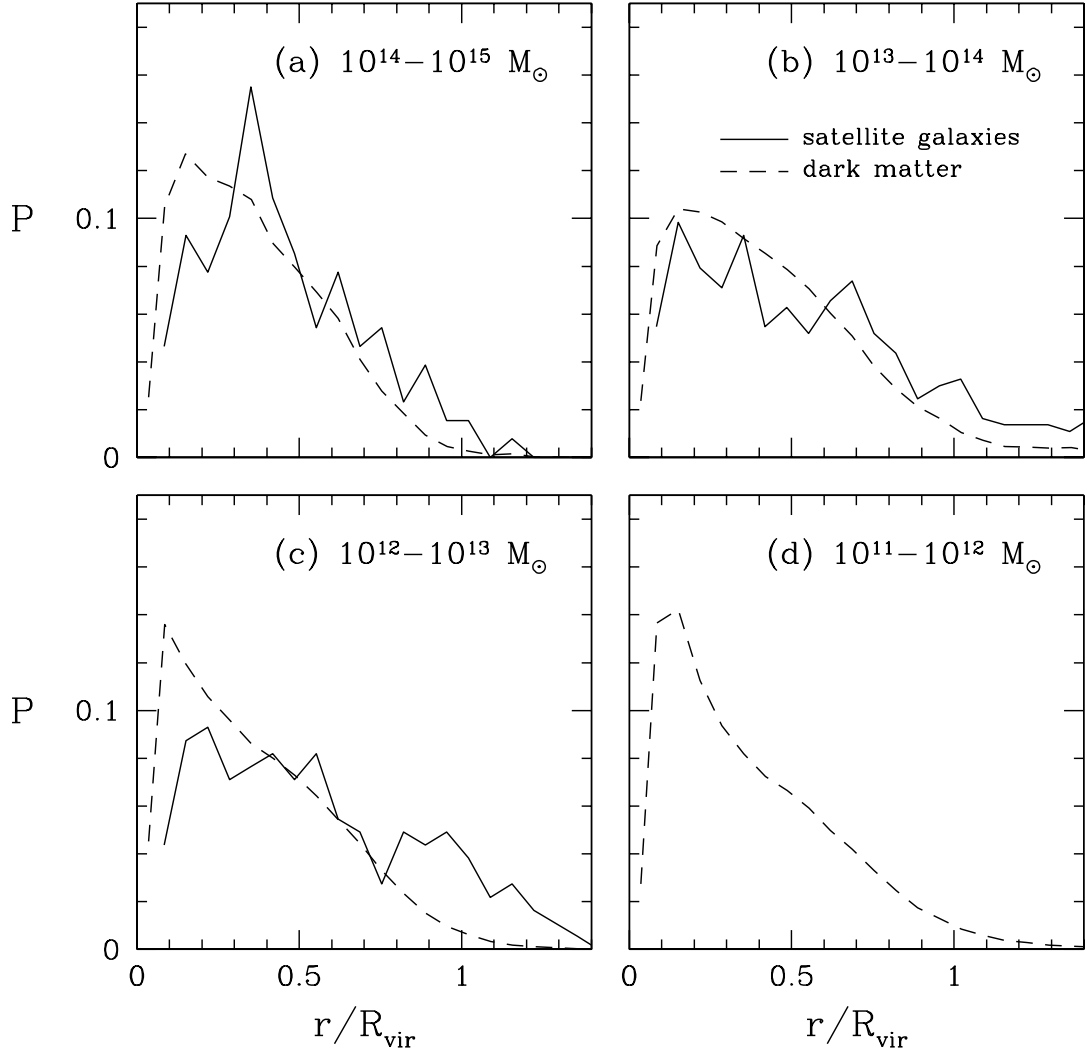}
\caption{Radial distribution of SPH ``satellite'' galaxies and dark matter 
within dark matter halos, in four halo mass bins.  Satellites are
all galaxies in each halo other than the central galaxy.  Galaxies and 
dark matter are represented by solid and dashed lines, respectively.
Halos in the lowest mass bin never have more than one galaxy, so by definition
they do not contain satellite galaxies.
} 
\label{fig:Nr}
\end{figure}

What about satellite (i.e., non-central) galaxies?  
Figure~\ref{fig:Nr} compares the radial profiles of satellite
galaxies (solid curves) to those of dark matter 
(dashed curves), in four halo mass bins.  Here we define profiles as
the fraction of objects in bins of $r/\Rvir$.
In all cases, the radial distribution of satellite galaxies traces
that of dark matter within halos fairly well.  
However, there is marginal evidence of a central core in the galaxy
distribution relative to that of the dark matter.
There are no satellite galaxies in the lowest halo mass 
bin (panel~d) because these halos never have more than one galaxy.

\begin{figure}
\plotone{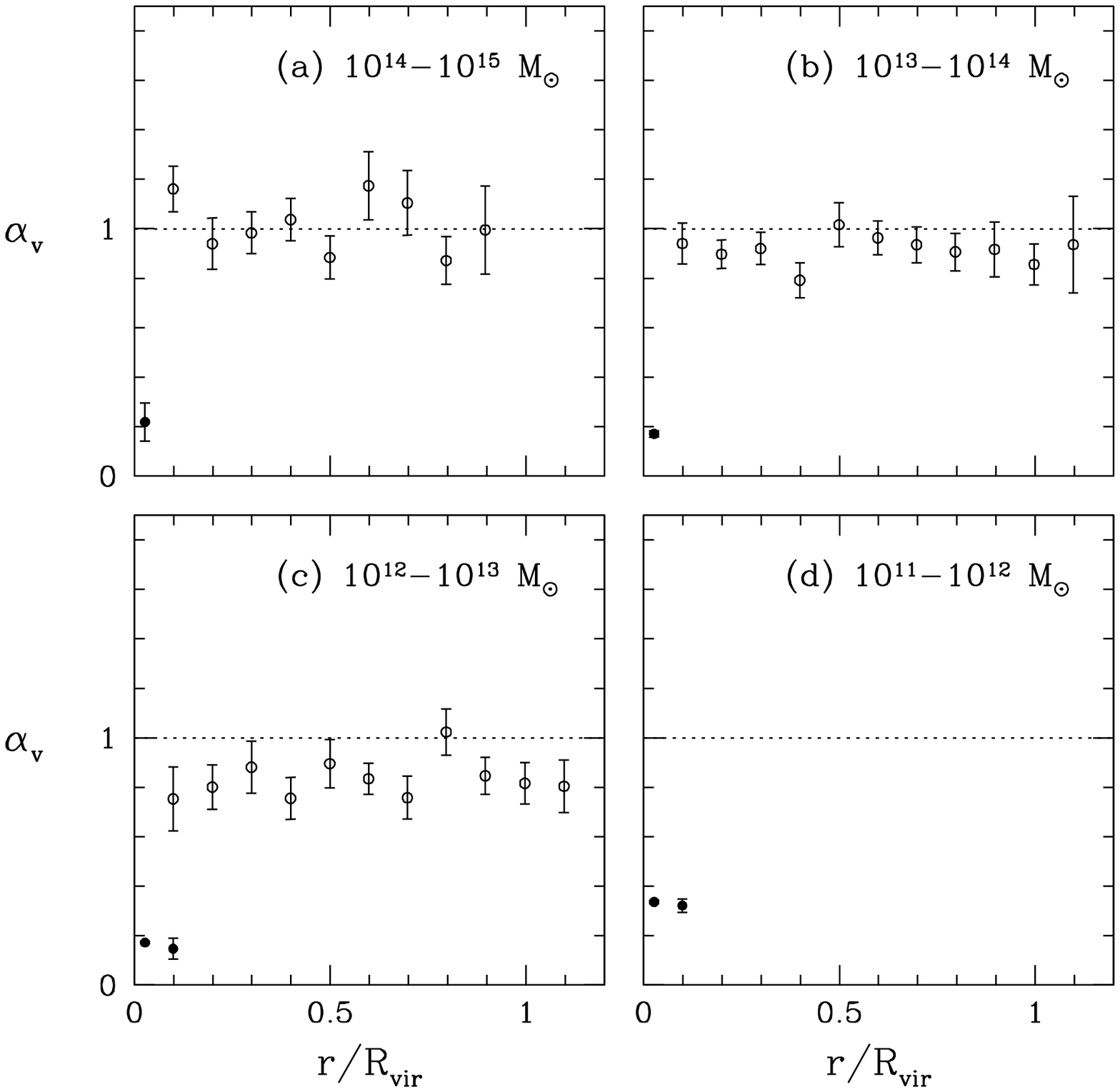}
\caption{Velocity bias factor 
$\alpha_v= \left<|\vg-\vh|\right>/\left<|\vm-\vh|\right>$
as a function of radius, in four halo mass bins.
Solid points show central galaxies and open points satellite galaxies,
with error bars showing uncertainty of the mean.
The value $\alpha_v=1$, corresponding to no 
velocity bias, is marked by the dotted line.  
} 
\label{fig:DVr}
\end{figure}

We now consider the relative velocities of galaxies and dark matter
within halos.  For every SPH galaxy, we measure $|\vg-\vh|$, 
the magnitude of its velocity relative to the halo center-of-mass velocity.  
We then average these measurements in bins of $r/\Rvir$, 
where $r$ is the galaxy's distance from the halo center (most bound particle).
We do the same for all dark matter particles.
Dividing these two functions gives the ``velocity bias'' parameter
$\alpha_v=\left<|\vg-\vh|\right>/\left<|\vm-\vh|\right>$ as
a function of radius within halos, similar to the $\alpha_v$ parameter 
defined by BW (which was assumed to be constant with radius).  
If galaxies have the same distribution of velocities as random dark
matter particles, then $\alpha_v=1$, while if
all galaxies move at their halo's mean velocity, then $\alpha_v=0$.
Figure~\ref{fig:DVr} shows $\alpha_v$ for central (solid points) and satellite
(open points) SPH galaxies as a function of $r/\Rvir$, in four halo mass bins.  
Error bars show the uncertainty in the mean $\alpha_v$.
Central galaxies have velocities that are substantially colder 
than the dark matter, as expected --- if they did not, then they would not 
remain close to the center.  Note that if the central galaxy is identified
with the most bound dark matter particle, its velocity should still be
set to the center-of-mass velocity; this difference can have an important
effect on velocity dispersion statistics \citep{benson00b}.
To a good approximation, satellite galaxies trace the dark matter velocity
distribution, though they exhibit a mild velocity bias $\alpha_v\sim 0.8-0.9$
in $10^{12}-10^{13}\Msun$ halos and $\alpha_v\sim 0.9-0.95$ 
in $10^{13}-10^{14}\Msun$ halos.

Overall, the SPH simulation supports a simple characterization of the galaxy 
distributions within halos: there is always one galaxy at the center of each 
halo (defined by the center of mass or, better, by the location of the most
bound dark matter particle) that moves at approximately the center-of-mass 
velocity, and any remaining satellite galaxies trace the spatial and velocity 
distribution of the dark matter.  This is just the characterization that 
has been used for most calculations of galaxy clustering from
SA/N-body models, and the only amendment that our SPH results suggest
is a mild velocity bias of satellites in intermediate mass halos and perhaps
a small central core in the galaxy distribution.  Here we have presented 
results for the whole $\ng=0.02\hvol$ sample, but we find that these results 
also hold for the lower space density samples.  However, when we split the
whole sample into two halves according to stellar population age, we find
that young satellite galaxies have a slightly bigger central core than old 
satellite galaxies.  We find no clear difference between the velocity bias of
young and old galaxies.


\section{``Central'' vs. ``Satellite'' Galaxies} \label{central}

\begin{figure}
\plotone{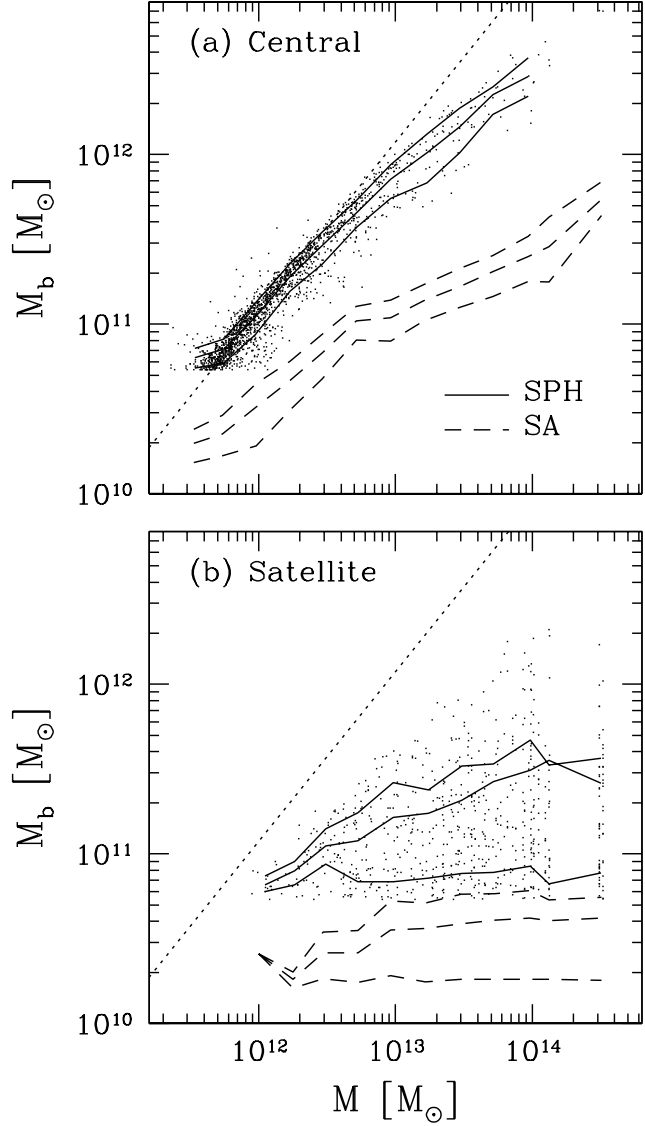}
\caption{Baryonic masses of (a) central and (b) satellite galaxies versus 
their 
host halo masses.  Each point represents the baryonic mass $M_b$ of an 
SPH galaxy. 
The middle solid curve shows the mean SPH $M_b$ in bins of halo mass, and  
the outer two solid curves enclose 60\% of all SPH central or satellite 
galaxies.  
The dashed curves show the same for (a) central and (b) satellite SA galaxies.
For comparison, dotted lines show the baryonic mass corresponding 
to the universal baryon fraction, $M_b=(\Omegab/\Omegam)M$.
} 
\label{fig:CenMass}
\end{figure}

Having established that SPH halos contain central galaxies, we can 
compare the properties of central and satellite galaxies in the 
SPH and SA models.
Figure~\ref{fig:CenMass} shows the distribution of baryonic masses for central 
(top panel) and satellite (bottom panel) galaxies.  Each point shows $M_b$ 
for an SPH galaxy versus its host halo mass.  The middle solid curves show the
mean $M_b$ computed in bins of halo mass, and the outer solid curves enclose
60\% of central/satellite SPH galaxies.  Dashed curves are similar to solid
ones, but for SA central (top panel) and satellite (bottom panel) galaxies.
Both models predict a tight correlation between central galaxy mass and 
host halo mass.  Satellite galaxy masses, on the other hand, are only 
weakly correlated with host halo masses, especially in the SA model.  
Furthermore, both models predict that
central galaxies are substantially more massive than their satellites and that
this difference increases with halo mass.  
This qualitative behavior is expected in a scenario where
the galaxies at the centers of large halos grow by accreting smaller
galaxies.  Central galaxies are also better positioned to accrete cooling
gas, since they sit at the maximum of the gas density profile --- in the
SA model, it is assumed that central galaxies accrete gas but 
satellite galaxies do not.
Our definition of SPH halo centers as the locations of the most bound 
dark matter particles tends to favor high masses for ``central'' galaxies
on its own, but Figure~\ref{fig:CenMass} does not look very different
if we use a center-of-mass definition instead, though in this case there
are some outliers at high $M$, which occur when substructure puts the center 
of mass closer to a low mass satellite.

As we have already seen in \S\ref{models:xi}, the SPH simulation predicts
much higher galaxy masses than the SA model.  For $M\la 3\times 10^{12}\Msun$, 
the baryonic mass of a typical SPH central galaxy is close to the halo mass 
multiplied by the universal baryon fraction (dotted line).  Some galaxies in 
this regime lie above the dotted line, indicating that they have accreted 
some gas from beyond the virial volume represented by the friends-of-friends
halo.  At high $M_b$, SPH galaxy masses are probably overestimated because 
of the resolution effects discussed in \S\ref{models:xi}.  Correcting for this 
effect would make little difference to central galaxies at low $M$,
but it would produce a sharper turnover in the trend of $M_b$ vs.\ $M$
at high $M$, making the gap between the SPH and SA predictions
roughly independent of halo mass.

\begin{figure}
\plotone{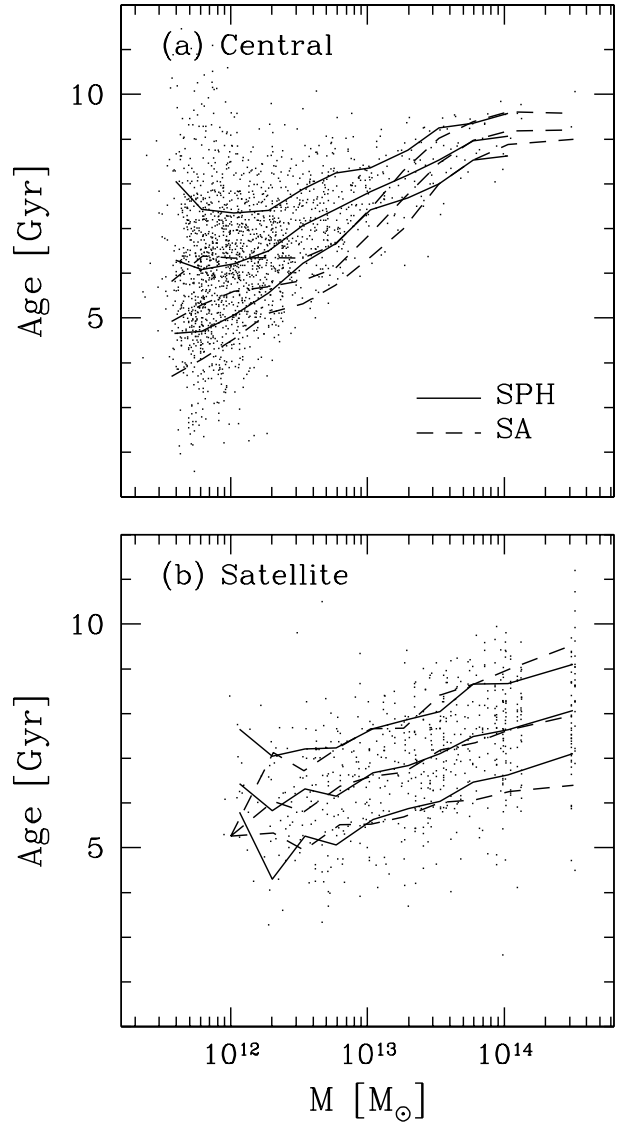}
\caption{Stellar ages of (a) central and (b) satellite galaxies versus their
host halo masses.  Each point represents the median mass-weighted stellar age
of an SPH galaxy. The middle solid curve shows the mean SPH galaxy age in bins 
of halo mass, and the outer two solid curves enclose 60\% of all SPH central or 
satellite galaxies.  The dashed curves show the same for the mean mass-weighted
stellar ages of (a) central and (b) satellite SA galaxies.
} 
\label{fig:CenAge}
\end{figure}
\begin{figure}
\plotone{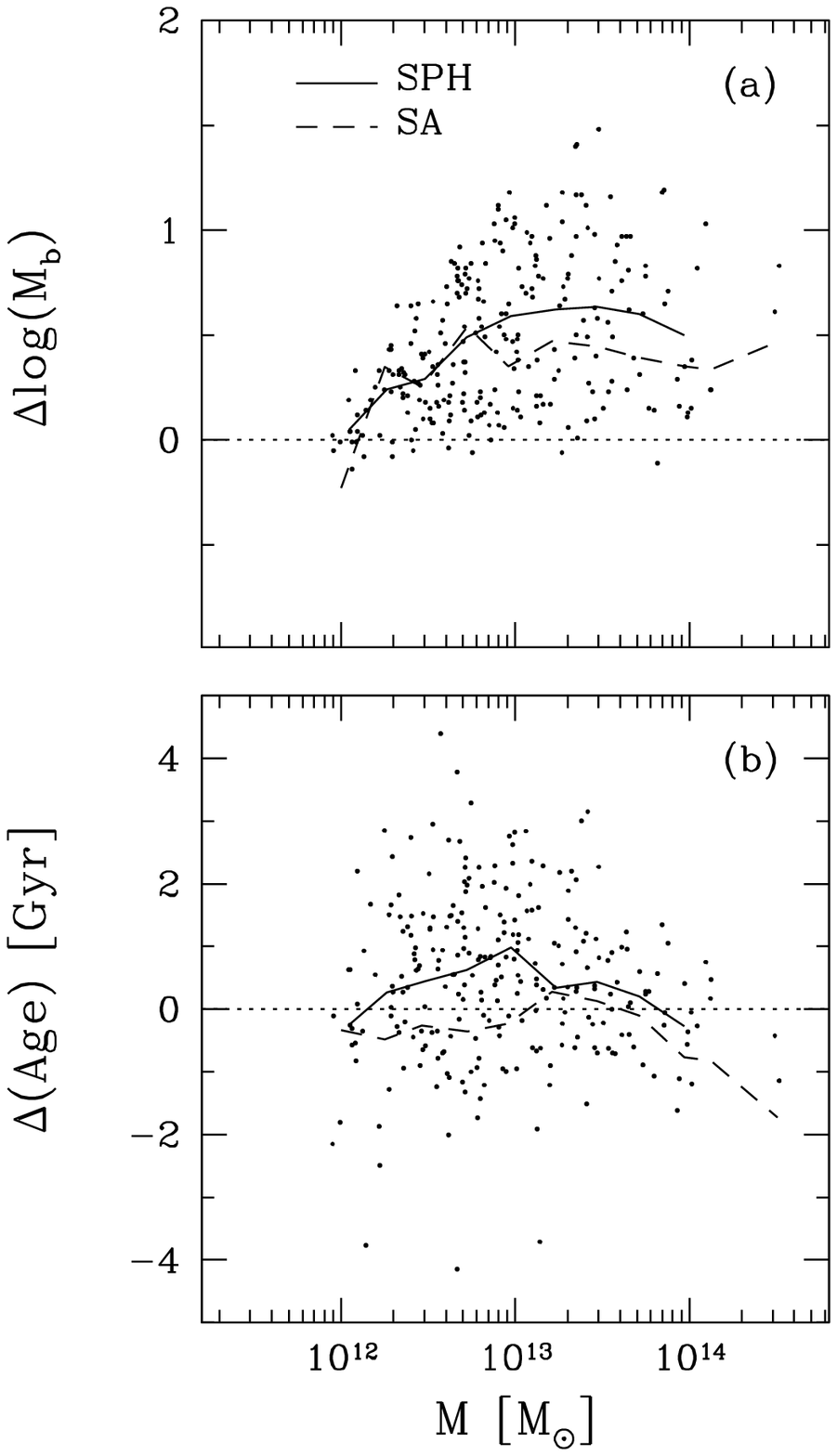}
\caption{(a) Difference in log$M_b$ between the
central galaxy of each halo and the most massive satellite galaxy.
Each point represents an SPH halo.  Solid and dashed curves show the
mean logarithmic mass difference for the SPH and SA halos, respectively.
(b) Like (a), but now showing the difference in age between the
central galaxy and the oldest satellite.
} 
\label{fig:DMDA}
\end{figure}

Figure~\ref{fig:CenAge} shows the distribution of stellar ages for central
(top panel) and satellite (bottom panel) galaxies, in a format similar
to Figure~\ref{fig:CenMass}.
Both models predict a clear correlation of central galaxy age with halo mass.  
Satellite galaxy ages also correlate with $M$, but not as strongly.  
As previously discussed in \S\ref{MassAge}, these trends reflect the 
earlier onset of structure formation in regions that become high mass
halos and the decline of gas accretion rates for galaxies that
reside in halos of high circular velocity.  The age scales in the SPH
and SA calculations agree reasonably well, though since they represent
median ages for SPH galaxies and mean ages for SA galaxies, we would not
expect perfect agreement in any case.

Figures~\ref{fig:CenMass} and~\ref{fig:CenAge} show that central
galaxies are generally more massive and older than average satellites
in the same halo.  Figure~\ref{fig:DMDA} addresses a related but distinct
question: is the central galaxy the most massive and oldest galaxy 
in the halo?  The top panel shows $\Delta\log M_b$, the logarithmic 
difference in mass between the central galaxy and the most massive 
non-central galaxy in the halo.  The bottom panel shows the age difference
between the central galaxy and the oldest non-central galaxy.
Points show individual SPH halos, and solid and dashed curves show
mean results in bins of halo mass for the SPH and SA models, respectively.
The central galaxy is consistently the most massive galaxy in the
halo, with both models predicting an average offset of $0.4-0.6\,$dex 
in halos with $M \ga 10^{13}\Msun$.  The decline in $\Delta\log M_b$
at lower $M$ is partly enforced by our baryonic mass threshold, since
halos that have only one
galaxy above the threshold (and ``satellites'' below it)
do not make it onto the plot.
The SPH central galaxies are frequently the oldest in the halo, but
the oldest satellite is only slightly younger on average, and sometimes
it is older.  The SA model predicts little mean age difference between
the central galaxy and the oldest satellite.  In the most massive 
halos, the central galaxy is younger than the oldest satellite, presumably
because it continues to accrete gas and form stars.
At a qualitative level, the predictions of both models are consistent
with the fact that observed central galaxies in massive groups and
clusters are usually luminous ellipticals or cD galaxies.


\section{Predictions for the Autocorrelation Function} \label{xiJen}

BW showed that, in the context of HOD models of galaxy bias, the
observed power-law form of the galaxy correlation function must
emerge from a delicate balance of several competing effects.
The low mass cutoff of $\N_M$, the high mass slope of $\N_M$,
the width of the $\PNN$ distribution, and the spatial distribution
of galaxies within halos each affect $\xi(r)$ in different ways.
A power-law $\xi(r)$ requires, at the least, a smooth connection
between the 1-halo and 2-halo contributions to the galaxy pair counts, 
and compatible logarithmic slopes in the 1-halo and 2-halo regimes.
Figure~\ref{fig:xi} shows that the HODs predicted by the SPH and SA
models achieve such an alignment, since both methods predict a correlation 
function that is roughly consistent with a power law for 
galaxies of space density $\ng=0.02\hvol$.  
However, since the emergence of the power law is somewhat fortuitous,
we expect that sufficiently precise predictions would show departures
from a pure power law shape.  These departures are hidden in 
Figure~\ref{fig:xi} by the relatively large error bars, which in turn
reflect the limited volume of our single, $50\hmpc$ simulation cube.

To improve the statistical precision of our $\xi(r)$ predictions, we 
bootstrap the HOD results derived from this $50\hmpc$ volume into
a larger volume N-body simulation.  Since $\PNM$ does not depend on the 
larger scale environment of halos (as shown in \S~\ref{env}), the HODs 
encode all the information needed to create new galaxy distributions 
that have the clustering properties predicted by the SPH and SA models.
We ran an N-body simulation, using version 1.1 of the parallel GADGET
code \citep{springel01}, with the same cosmological parameters and 
initial power spectrum used for the SPH and SA models (described 
in \S~\ref{models}).  The simulation follows the evolution of $256^3$ 
dark matter particles in a $112.5\hmpc$ box, with a particle mass about 
twice that of dark matter particles in the SPH simulation.  The 
gravitational softening length is $15\hkpc$.  We identify dark matter 
halos using a friends-of-friends algorithm with a linking length of 
0.173 times the mean inter-particle separation (exactly as we did for the
SPH simulation).  We populate each halo with a number of galaxies drawn from
the $\PNM$ distributions predicted by the SPH and SA models.  In the high
halo mass regime, the SPH model makes no direct prediction of $\PNM$ because 
the simulation volume does not contain enough high mass halos, and we have
not computed SA predictions for halo masses not represented in the
simulation.  In this regime, therefore, we use the fitting formula of 
equation~(\ref{eqn:Nfit}) with the parameter values in Table~1 to compute 
$\N$, and we draw a number of galaxies assuming a Poisson $\PNN$ distribution.  
The latter assumption is reasonable because we have shown that
both models predict Poisson pair counts $\NN_M$ in high mass halos
(Fig.~\ref{fig:NN}).  Once we have determined $N$ for a given halo, we place the 
first galaxy at the center of mass of the halo and any remaining galaxies
at the locations of random dark matter particles within the halo.  Once again,
these are reasonable assumptions given our results on the spatial distribution 
of galaxies within halos in \S~\ref{profile}.  In this way, we create 
galaxy distributions in the $112.5\hmpc$ box using the HODs predicted by the
SPH and SA models for galaxies of space densities $\ng=0.02, 0.01$, and 
$0.005\hvol$.

\begin{figure}
\plotone{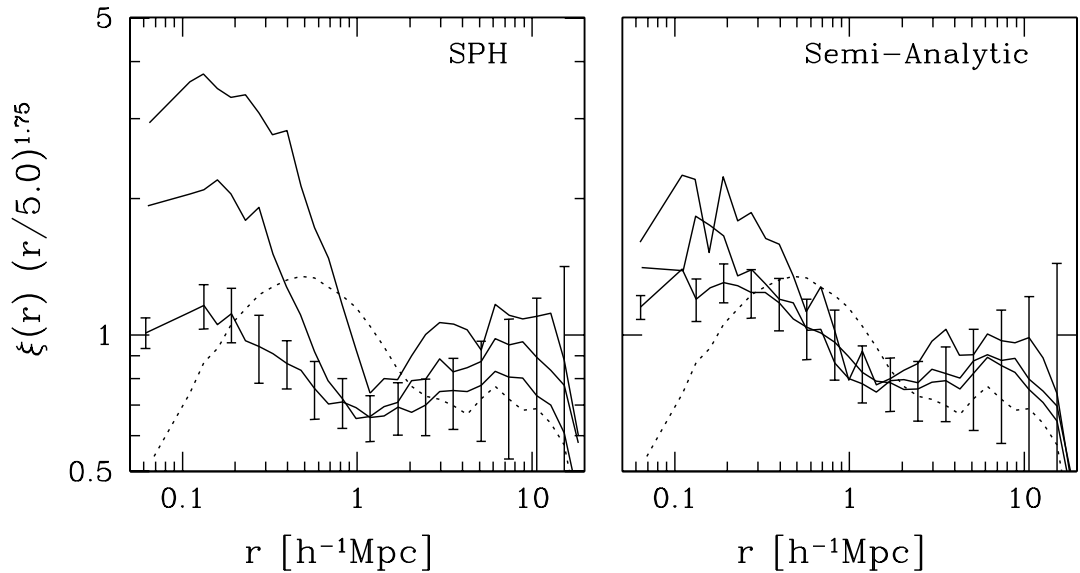}
\caption{Two-point correlation functions, divided by the power law 
$(r/5.0)^{-1.75}$, computed by applying the predicted SPH (left panel) and 
SA (right panel) $\PNM$ to a larger N-body simulation of the same cosmological 
model.  For each model, the three correlation functions 
correspond to galaxy populations of space densities $\ng=0.005\hvol$ 
(top lines), 
$0.01\hvol$ (middle lines), and $0.02\hvol$ (bottom lines).  Errors in the mean 
estimated from jackknife resampling using the eight octants of the cube are 
shown for the $\ng=0.02\hvol$ samples.  Also shown is the dark matter 
correlation function of the N-body simulation (dotted line).
}
\label{fig:xiJen}
\end{figure}

Figure~\ref{fig:xiJen} shows the correlation functions for these bootstrapped
galaxy distributions.  To investigate departures from a power-law
shape, we have divided $\xi(r)$ by the power law $(r/5.0)^{-1.75}$,
and we use logarithmic axes so that power laws of other slopes would
still be straight lines in the plot.  Errors in the mean estimated from 
jackknife resampling using the eight octants of the cube are shown for 
the $\ng=0.02\hvol$ samples.  The errors are highly correlated, as 
one can see from the smoothness of the curves relative to the size of 
the error bars.  Nonetheless, these error bars are small enough to reveal 
interesting features that were not evident before.  In all cases, 
$\xi(r)$ dips below a power-law extrapolation at $r\sim 1\hmpc$, then 
rises more steeply from $r\sim 1$ to $r\sim 0.2\hmpc$ before returning 
to the original power law at still smaller scales.  The inflection at 
$r\sim 1\hmpc$ arises at the transition between the 2-halo term, which 
flattens (toward smaller $r$) and eventually cuts off on the scales
of halo virial radii, and the 1-halo term, which rises steeply for $r$
close to the virial radii of the most massive halos (see, e.g., BW Fig.~7).
The predicted departures are stronger for more massive galaxy samples
with lower space densities, and Figure~\ref{fig:xiJen} shows that 
even the small difference between the SPH and SA HODs is enough to 
have a significant quantitative impact, with the SPH model predicting
stronger departures from a power law.  For the high space density sample,
the difference between the SPH and SA correlation functions is caused 
mainly by the difference in the high mass slopes of $\N_M$, but for the 
low space density samples, differences in the cutoff and plateau regime 
dominate.  We have also investigated the impact on the predicted $\xi(r)$ 
of assuming a nearest-integer distribution for all halo masses instead of using
the true predicted $\PNN$.  We find that assuming a Nint distribution at all 
masses leads to an underestimate of $\xi(r)$ by $10-20\%$ at $r \la 0.5$ for 
$\ng=0.02\hvol$, growing to $30-40\%$ for the lower space densities.

\begin{figure}
\plotone{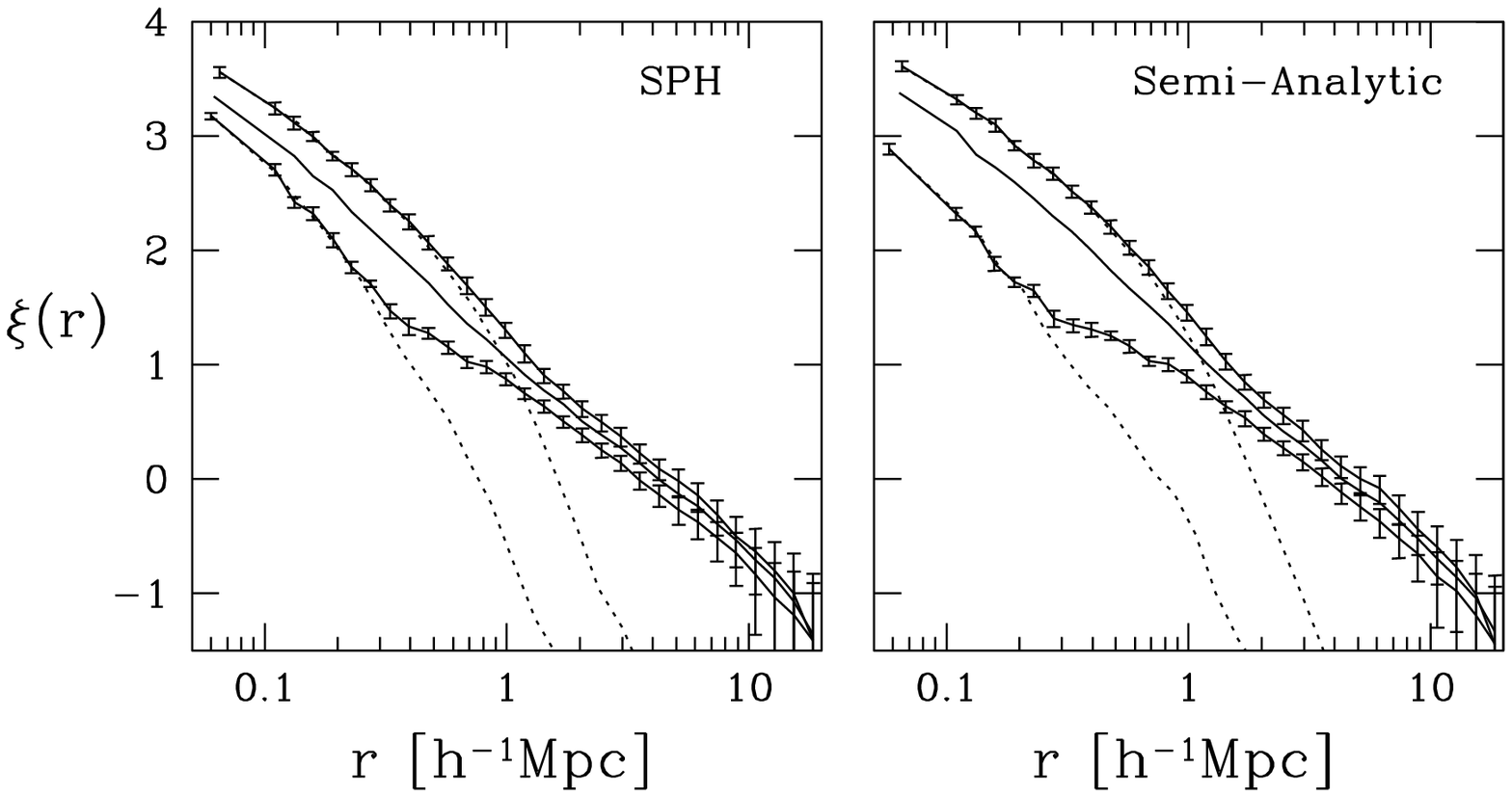}
\caption{Two-point correlation functions of old (top curves), young (bottom curves),
and all (middle curves) galaxies in the SPH (left panel) and SA (right panel) models.
Correlation functions were computed by applying the predicted $\PNM$ for each case
to a larger N-body simulation of the same cosmological model.  
Errors in the mean that were estimated from jackknife resampling using the eight 
octants of the cube are shown for the young and old halves.  Also shown are the 
1-halo terms (dotted curves) for the young and old halves.
}
\label{fig:xiJen2}
\end{figure}

Figure~\ref{fig:xiJen2} shows predicted correlation functions for the old and
young halves of the SPH and SA $\ng=0.02\hvol$ samples.  As before, these were 
computed by populating the larger volume N-body simulation with galaxies using
the predicted $\PNM$ for old and young galaxies (of which the mean occupations
are seen in Fig.~\ref{fig:Nage2}).  In the high halo mass regime, where we do 
not have secure calculations of $\PNM$ because of our limited number of halos,
we compute $\N$ by extrapolation, and we draw a number of galaxies assuming a 
Poisson $\PNN$ distribution.  Figure~\ref{fig:xiJen2} also shows the 1-halo 
terms for old and young galaxies.  The most striking difference between old and 
young galaxies is that the 1-halo term for young galaxies is severely depressed 
in amplitude compared to that for old galaxies.  This depression is caused by a 
very high fraction of young galaxies that are isolated in their own halos.  As 
a consequence, the 1-halo and 2-halo terms for young galaxies meet, and thus 
create a feature in $\xig$, at a much smaller scale ($r\sim0.3\hmpc$) than for 
old galaxies.  The depression in the young 1-halo term predicted by the SA model 
is stronger than the SPH one, due to the lower fraction of young galaxies in 
high multiplicity halos seen in Figure~\ref{fig:Nage2}.

Most observational measurements of the galaxy correlation function have 
appeared consistent with a power law on scales $r \la 5\hmpc$ (e.g., 
\citealt{norberg02,zehavi02}; and numerous references therein), though 
there have been suggestions of a ``shoulder'' or flattening in the range 
$5\hmpc \la r \la 10\hmpc$ 
(e.g., \citealt{dekel84,baugh96,gaztanaga01,hawkins03}).
The theoretical predictions (here and in earlier papers) suggest that
departures from a pure power law should be measurable at sufficiently
high precision, with the generic expectation being an inflection in
$\xi(r)$ near the scale of the 1-halo to 2-halo transition.  Motivated in 
part by these predictions, \cite{zehavi03} have recently examined the 
projected correlation function of luminous $(M_r<-21)$ galaxies from the SDSS.  
They find a statistically significant departure from a power law of just this 
predicted form.  The space density of the Zehavi et al.\ sample is lower than 
that of our highest mass-threshold sample, so we cannot directly test the 
results shown in Figure~\ref{fig:xiJen}, but their measurements can be well 
fit by an HOD model that is qualitatively similar to that found here.  
\cite{magliocchetti03} have shown that the projected correlation functions of 
2dFGRS galaxies can also be fit well by HOD models.  As Figures~\ref{fig:xiJen}
and~\ref{fig:xiJen2} show, the SPH and SA models make a number of distinctive 
predictions about the dependence of the correlation function amplitude and 
shape on galaxy mass and age, and these predictions can be tested in the
near future.


\section{Summary and Discussion} \label{conclusions}

The SPH and SA predictions of the galaxy HOD agree remarkably well.
The qualitative features of these predictions can, for the most part,
be readily interpreted in terms of galaxy formation physics.  The mean 
occupation function for galaxies above a mass threshold has a cutoff, a 
slowly rising plateau, and a steeper ``high occupancy'' regime.  In the 
SPH simulation, the cutoff is determined by the universal baryon fraction.  
In the SA model, the cutoff occurs at a substantially higher mass than 
the universal baryon fraction requires, corresponding to only $\sim$25\% 
of the gas in low mass halos remaining in galaxies.  This is due to the
feedback mechanism, which in the SA model is tuned to produce the observed
galaxy luminosity function.  The cutoff is nearly as sharp in the SA model 
as in the SPH simulation because feedback is tightly regulated by halo mass, 
with no scatter that allows some low mass halos to retain substantially more than 
$\sim$25\% of their baryons.  In the low occupancy regime, $\N$ rises slowly 
with $M$ because additional mass tends to go into making more massive single 
galaxies instead of multiple low mass galaxies.  The logarithmic slope of $\N_M$ 
steepens for $\N \gtrsim 2$, presumably representing a transition to a regime in 
which the galaxies in a halo were formed mostly in the lower mass progenitor 
halos that merged to create it.  Even in this regime, the slope is below unity, 
since the overall efficiency of galaxy formation (fraction of mass in galaxies) 
is lower in high mass halos, probably a consequence of longer cooling times
and less efficient filamentary ``channeling'' of gas into galaxies.
For higher baryonic mass thresholds, the mean occupation function 
shifts horizontally along the $\log M$ axis, so this physical explanation of 
its form is not sensitive to the particular threshold adopted.

Since the considerations that account for the form of $\N_M$ are quite general, 
it is perhaps not surprising that the SPH and SA calculations agree.  More 
impressive is the agreement on fluctuations about the mean occupation.  Both 
calculations predict pair counts well below the expectation for Poisson 
fluctuations at low mass, close to those of the minimal fluctuation, 
nearest-integer model.  As many authors have emphasized 
(\citealt{benson00,seljak00,peacock00,scoccimarro01}; BW),
these sub-Poisson fluctuations are 
essential in understanding the power-law form of the observed 
galaxy correlation function.  Triple counts are also substantially sub-Poisson, 
though not minimal, and the two calculations again agree.  
The explanation of sub-Poisson fluctuations is not trivial, but it clearly 
has to do with statistics of halo merger histories and the timescales of 
galaxy mergers following halo mergers.  In particular, near equal-mass halo 
mergers are rare enough that one must go substantially above the cutoff mass 
$\Mmin$ (by a factor $\sim10$) before one is likely to get two galaxies above 
the mass threshold in a single halo, instead of one, more massive, galaxy.
The probability of empty halos follows the nearest-integer prediction,
$P(0|M)={\rm max}(1-\N_M,0)$, almost exactly in both models.
Because the cutoff of $\N_M$ is somewhat sharper in the SPH calculation,
the transition from occupied halos to empty halos (at the specified
baryonic mass threshold) is also somewhat sharper, occurring over about
a factor of two in halo mass.  For masses above the cutoff, the probability
that a halo is empty is much lower than the Poisson expectation
$\exp(-\N_M)$.  In this regime, strongly sub-Poisson statistics
indicate the regularity of the formation process in single-galaxy
halos, which produces a tight correlation between galaxy mass and halo mass.

The mean occupation is very different for galaxies in different age quartiles, 
at both the low and high mass ends of $\N_M$.
Young galaxies are rare in high mass halos and vice versa.
This difference reflects a 
tendency for star formation to begin early in regions that will become part 
of high mass halos and for star formation to shut off when a galaxy falls 
into a halo of larger virial mass,
or to slow down when the halo size or virial temperature becomes too large.
The qualitative agreement between the SPH and SA calculations is again not 
surprising, but the degree of quantitative agreement is impressive.

The SPH simulation also shows that the mean occupation function is 
independent of the large scale environment of the halo, extending to the 
galaxy regime the result that \citet{lemson99} found for the merger histories 
of dark matter halos.  This is 
a fundamental and encouraging result, supporting the assumption made in 
merger-tree based SA methods, 
and supporting the claim that the HOD provides a statistically 
complete description of galaxy bias.

A detailed analysis of the SPH simulation also 
supports simple assumptions about the 
galaxy distribution within halos.  Most halos have a galaxy near the center of 
mass moving at close to the center-of-mass velocity.  
The central galaxy is almost always the most massive galaxy in the halo,
and it is usually among the oldest galaxies in the halo.
The remaining, satellite galaxies 
have the same spatial and velocity distribution as the dark matter, 
except for a small ($\alpha_v \sim 0.8-0.95$)
velocity bias in intermediate mass halos.
These assumptions are thus reasonable to 
use when making galaxy clustering predictions, and they are the ones usually 
incorporated in studies that combine N-body and semi-analytic methods 
(\citealt{kauffmann99}; \citealt{benson00}; \citealt{somerville01}).  
In the long 
term, these predictions can be tested empirically using clustering and
galaxy-galaxy lensing data of the sort provided by 2dFGRS and SDSS.  

As we have emphasized in separate papers on the SA model and SPH simulation
(\citealt{benson00}; \citealt{weinberg02a}), the complex relation between 
galaxies and mass is crucial in reproducing the observed nearly power-law 
form of the galaxy correlation function, since the dark matter $\xi(r)$ is not 
a power law in CDM-type models.  Since the emergence of the power law is to 
some extent a coincidence, the HOD framework generically predicts that there 
should be departures from a power-law $\xi(r)$ once it is measured with 
sufficient precision, especially if one considers different subclasses of 
galaxies.  We have tried to predict these departures by bootstrapping our 
results into a larger volume N-body simulation.  
The small differences between the SA and SPH $\PNM$ translate into noticeable 
differences in the predicted $\xi(r)$, and the accuracy of the SPH prediction 
is still limited to some extent by the small number of high mass halos in the 
SPH volume.  However, improved statistics show that both models predict
an inflection in $\xi(r)$ at $r\sim 1\hmpc$, with the departures from a power 
law being stronger for the SPH model than for the SA model and stronger for 
more massive galaxies in both cases.  Recent observational analyses provide 
evidence for just such a feature in the correlation function of luminous 
galaxies \citep{zehavi03}.

The most surprising aspect of our results is that the SPH and SA models give 
such similar predictions for $\PNM$ despite having very different galaxy mass 
functions.
The different mass functions reflect the different treatments of cooling in the 
SA and SPH calculations (including the geometric idealizations in the former 
and the numerical resolution limitations in the latter), and the greater 
importance of stellar feedback in the SA calculation.  The similarity of 
the predicted halo occupations implies that the HOD is
determined by physics that is robust to these cooling and feedback
differences.  Obviously it is crucial 
that we select galaxy samples based on number density, using only the rank 
order of the baryonic masses rather than their absolute values; for a fixed 
mass threshold, 
the numbers of galaxies would be different in the two calculations,
and the HODs 
could not possibly match.  Similarity of HODs suggests that the rank order of 
galaxy masses and the rank order of stellar population ages are related in 
fairly simple ways to the merger histories of the dark matter halos, which 
should be statistically similar in the two methods.  
For example, the baryonic mass of a galaxy 
could be monotonically related 
(by non-linear functions that are different in the 
two calculations) to the mass of its parent halo at the last time this 
galaxy was ``central.'' 
The age could be determined (at least in a rank-order sense) by the 
formation time of this halo and the time at which its central object's
star formation is suppressed by falling into a larger halo and becoming a 
satellite.  Mergers within halos also affect $\PNM$, but this process
should be similar in the two calculations, since the assumptions about mergers 
in the SA model reproduce the results of numerical calculations \citep{benson02}.

We can test this idea --- that the form of the HOD is driven largely
by dark matter dynamics --- by re-analyzing the SA galaxy population
without using the predicted baryonic properties.
For every SA galaxy, we identify its 
parent halo at the last time this galaxy was central,
and we denote the halo mass at that time by $\Mmax$
and the time itself by $\tmax$.  
We then create galaxy samples with the same space densities
as before but use thresholds in $\Mmax$ rather than in baryonic mass.  
Dotted curves in Figures~\ref{fig:Nn} and~\ref{fig:NNn} show
the first and second moments of $N(M)$, respectively,
for these samples defined by $\Mmax$ thresholds.
It is interesting both that this simple mapping works 
as well as it does and that it does not work perfectly.
The qualitative similarity of the mean occupations supports the idea that
$\PNM$ is largely governed by robust physics connected to halo merger
histories.  However, obvious differences, 
in particular the steeper slope of $\N_M$ at the high mass end, 
indicate that additional physical processes must also play a significant role.
It is encouraging that the SPH and SA models agree with each 
other better than either model agrees with the ``rank by $\Mmax$'' calculation.
The fact that they predict fewer galaxies in high mass halos suggests
that galaxy accretion rates begin to drop in overdense environments even 
before their parent halos fall into larger halos, allowing some of these
galaxies to drop below the baryonic mass threshold and be replaced
(in a sample of fixed space density) by galaxies forming in lower mass halos.
Recent studies provide observational support for this effect, showing
decreased star formation rates in galaxies that are well outside the
virial radii of rich clusters (e.g., \citealt{lewis02,gomez03}).

\begin{figure}
\plotone{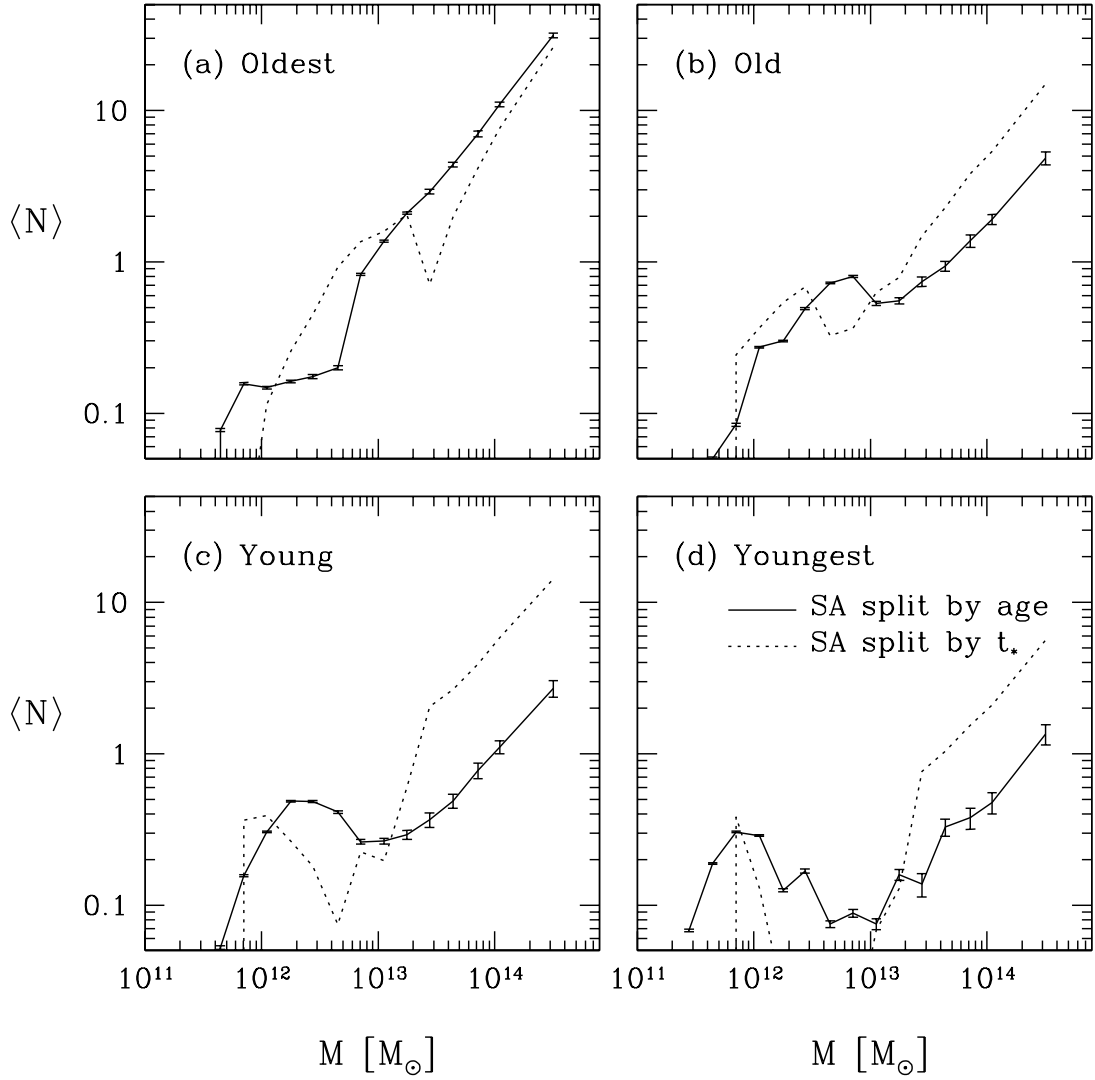}
\caption{Mean occupation $\N_M$ of SA galaxies, for quartiles
of mean stellar galaxy age (solid curves) and merger tree ``age'' parameter 
$t_*$ (dashed curves, defined in \S~\ref{conclusions}).  
}
\label{fig:tstar}
\end{figure}

We can map the stellar age of a galaxy onto the halo merger tree in a similar
fashion.  We make the simple assumption that the star formation in each galaxy
starts at a time $\tstart$ when its parent halo first exceeds
a mass $10^{11}\Msun$, a factor $\sim 4$ below the cutoff for 
$\ng=0.02\hvol$.  We assume that the star formation ends at a time $\tstop$ 
when the galaxy becomes a satellite ($\tstop=\tmax$) or when the cooling time 
in its parent halo becomes too long ($\tstop=t_{\rm hot}$) --- whichever 
happens first.  We identify $t_{\rm hot}$ as the time when the parent halo mass 
exceeds $10^{13}\Msun$.  Finally, we assume a constant star formation rate 
between $\tstart$ and $\tstop$, making the mean stellar age of the galaxy 
$t_* = t_0 - (\tstop-\tstart)/2$.  Figure~\ref{fig:tstar} compares the mean 
occupation functions for SA galaxies split into quartiles according to $t_*$ 
to those of quartiles defined by true stellar age (repeated from 
Figure~\ref{fig:Nage}).  Our highly simplified approximation yields the right 
qualitative dependence of $\N_M$ on galaxy age, but it differs substantially 
in the quantitative details.  The $\Mmax$ and $t_*$ models still make indirect 
use of the baryonic physics in the SA calculation, since baryonic masses 
affect which galaxies remain as distinct entities in larger halos and which
are destroyed by dynamical friction.  Results from a more concerted effort to 
model the HOD entirely with dark matter dynamics will be reported elsewhere 
(Berlind, Bullock, Kravtsov, Wechsler, and Zentner in preparation).

Finite resolution of the SPH simulation has limited our comparison 
of predictions to relatively massive galaxies, and to age rather 
than morphology as a distinguishing characteristic of sub-classes.
Insensitivity of predictions to calculational details is likely to 
change when one gets to low mass galaxies or to characteristics that 
are determined by interactions with environment.  For example, we might 
expect the HOD of low mass galaxies to be 
quite different if their luminosity is controlled by photoionization 
or by supernova feedback.  Likewise, models that ascribe bulge-to-disk 
ratios mainly to merger histories, to secular evolution of disks, to weak 
perturbations in group/cluster environments, or to interaction with the IGM
could yield quite different 
predictions for the $\PNM$ of different morphological types.  
The process of inferring halo occupations from observed galaxy clustering
has already begun, and it should accelerate over the next few years with
improved measurements from the 2dFGRS and the SDSS.
Empirical HOD determinations for galaxies with $L\ga L_*$,
classified by luminosity and color or spectral type,
will test the basic predictions of the current scenario of galaxy 
formation, as presented here.  They 
will also sharpen the constraints on cosmological models by
removing bias as a degree of freedom in matching observed clustering.
Empirical determinations of HODs for faint galaxies and for morphological 
subtypes should yield insight into the physics that governs the low
end of the luminosity function and the origin of galaxy morphology.

\acknowledgments 

This work was supported by NSF Grants PHY-0079251, AST-0098584, 
and AST-9802568.  DHW acknowledges the hospitality of the 
Institut d'Astrophysique de Paris and the support of the French CNRS
during phases of this work.

\newpage
\begin{center}
\centerline{\small Table~1. Fit parameters for $\N_M$ fitting function.}
\smallskip
\begin{tabular}{lcccccccccc}
\tableline
\tableline
\multicolumn{1}{l}{$\ng$} &
\multicolumn{1}{c}{$\Mbmin$} &
\multicolumn{1}{c}{$\Mmin$} &
\multicolumn{1}{c}{$\nu$} &
\multicolumn{1}{c}{$\Mcrit$} &
\multicolumn{1}{c}{$\mu$} &
\multicolumn{1}{c}{$\alpha$} &
\multicolumn{1}{c}{$\beta$} &
\multicolumn{1}{c}{$M_1$} &
\multicolumn{1}{c}{$\N(\Mcrit)$} &
\multicolumn{1}{c}{$\Delta$(log$\N)_{\mathrm{max}}$} \\
\tableline
SPH & & & & & & &\\
0.02  & 10.74 & 11.70 & 6.2 & 12.70 & 1.6 & 0.09 & 0.74 & 12.00 & 1.5 & 0.06\\
0.01  & 11.10 & 12.05 & 8.2 & 12.85 & 5.0 & 0.21 & 0.56 & 12.30 & 1.4 & 0.07\\
0.005 & 11.38 & 12.35 & 7.8 & 13.40 & 1.9 & 0.30 & 0.70 & 12.75 & 1.8 & 0.14\\
\tableline
SA & & & & & & &\\
0.02  & 10.16 & 11.75 & 2.9 & 13.00 & 1.8 & 0.22 & 0.91 & 12.15 & 2.0 & 0.09\\
0.01  & 10.50 & 12.05 & 2.2 & 13.15 & 1.8 & 0.20 & 0.81 & 12.60 & 1.6 & 0.07\\
0.005 & 10.75 & 12.35 & 2.5 & 13.60 & 2.0 & 0.27 & 0.81 & 12.90 & 1.8 & 0.06\\
\tableline
\label{tab:1}
\end{tabular}
\end{center}
\center Note---Mass columns list log$_{10}(M/\Msun)$.

\end{document}